\documentclass{aa}

\usepackage[T1]{fontenc}
\usepackage{ae,aecompl}

\usepackage{graphicx}
\usepackage{txfonts}
\usepackage{amssymb}
\usepackage{mathtools}
\usepackage{comment}
\usepackage{subfig}
\usepackage{hyperref}
\hypersetup{
    colorlinks = true,
    citecolor = blue,
}

\graphicspath{ {./images/} }

\defcitealias{soldateschi_2020}{SBD20}
\defcitealias{soldateschi_2021}{SBD21}

\begin{document}

\title{Quasi-universality of the magnetic deformation of neutron stars in general relativity and beyond}
\titlerunning{Magnetic deformation of NSs}

\author{J. Soldateschi
          \inst{1,2,3}
          \and
          N. Bucciantini\inst{2,1,3}
          \and
          L. Del Zanna\inst{1,2,3}
          }

        \offprints{J. Soldateschi, \email{jacopo.soldateschi@inaf.it} or N. Bucciantini, \email{niccolo.bucciantini@inaf.it}}

   \institute{Dipartimento di Fisica e Astronomia, Universit\`a degli Studi di Firenze, Via G. Sansone 1, I-50019 Sesto F. no (Firenze), Italy
         \and
             INAF - Osservatorio Astrofisico di Arcetri, Largo E. Fermi 5, I-50125 Firenze, Italy
                \and
                        INFN - Sezione di Firenze, Via G. Sansone 1, I-50019 Sesto F. no (Firenze), Italy
             }

\date{Received XXX; accepted YYY}

\abstract{
Neutron stars are known to  host extremely powerful magnetic fields. Among its effects, one of the consequences of harbouring such fields is the deformation of the neutron star structure, leading, together with rotation, to the emission of continuous gravitational waves. On the one hand, the details of their internal magnetic fields are mostly unknown. Likewise, their internal structure, encoded by the equation of state, is highly uncertain. Here, we present a study of axisymmetric  models of isolated magnetised neutron stars  for various realistic equations of state considered viable by observations and nuclear physics constraints. We show that it is possible to find simple relations between the magnetic deformation of a neutron star, its Komar mass, and its circumferential radius in the case of purely poloidal and purely toroidal magnetic configurations that satisfy the criterion for equilibrium in the Bernoulli formalism. Such relations are quasi-universal, meaning that they are mostly independent from the equation of state of the neutron star. Thanks to their formulation in terms of potentially observable quantities, as we discuss, our results could help to constrain the magnetic properties of the neutron star interior and to better assess the detectability of continuous gravitational waves by isolated neutron stars, without knowing their equation of state.
Our results are derived both in general relativity and in scalar-tensor theories (one of the most promising extensions of general relativity), in this case by also considering the scalar charge. We show that even in this case, general relations that account for deviations from general relativity still hold, which could potentially be used to set constraints on the gravitational theory.
}


\maketitle

%

\section{Introduction}\label{sec:intro}
Neutron stars (NSs) are the most compact material objects in the  Universe and are known to harbour extremely powerful magnetic fields. Among the variety of observed NSs, a sub-class of them hosts the most powerful magnetic fields known to us: magnetars {\citep{duncan_1992,thompson_1993,thompson_1995,thompson_1996}}, whose name refers to NSs that were shown to {exhibit energetic bursting (soft gamma repeaters) and persistent (anomalous X-ray pulsars) activity} {\citep{kouveliotou_1998,gavriil_2002,mereghetti_2015}}. In fact, while the surface magnetic field of NSs has been inferred to be in the range of $10^{8-12}$G for radio and $\gamma$-ray pulsars \citep{asseo_2002,spruit_source_2009,ferrario_magnetic_2015}, estimates for magnetars have reached $10^{15}$G \citep{olausen_2014,popov_origins_2016}. However, magnetars represent a small subset of NSs: to this day, the known population of magnetars consists of just over 30 sources \citep{olausen_2014}\footnote{See the catalogue website for an up-to-date list of known magnetars: \url{http://www.physics.mcgill.ca/~pulsar/magnetar/main.html}.}, compared to a few thousand regular pulsars \citep{atnf_2005}\footnote{See the catalogue website for an up-to-date list of known pulsars: \url{https://www.atnf.csiro.au/research/pulsar/psrcat/}.}; nonetheless,  it is believed that they might represent a significant fraction of the young NS population \citep{kaspi_2017}.
\\\\
As opposed to the surface and magnetospheric magnetic field on NSs, which can be probed and constrained with some accuracy through a variety of different methods \citep{rea_2010,guver_2011,rea_2012,kontorovich_2015,rodriguez_2016,jankowski_2017,staubert_2019}, the geometry and strength of their internal magnetic fields remain largely unknown. It has been predicted that it may reach values as high as $10^{16}$G inside magnetars and up to $10^{17-18}$G in newly born proto-NSs \citep{del_zanna_chiral_2018,ciolfi_2019,franceschetti_2020}. Indeed, the possibility of highly magnetised and rapidly rotating proto-NSs is at the foundation of the so-called millisecond magnetar model for long and short gamma-ray bursts {\citep{usov_1992,Metzger_Giannios+11a,rowlinson_2013}}; moreover, it has been suggested that these objects are a possible source of fast radio bursts {\citep{lyubarsky_2014,beloborodov_2017,metzger_2019,hessels_2018,platts_2019,dallosso_2021}}. What appears less clear is the geometry of their internal magnetic field. While it is widely known that neither purely poloidal nor purely toroidal  configurations are stable {\citep{prendergast_equilibrium_1956,chandrasekhar_1956,chandrasekhar_1957,tayler_adiabatic_1973,markey_1973,markey_1974,tayler_1980,bocquet_rotating_1995,oron_relativistic_2002,braithwaite_stable_2006,braithwaite_evolution_2006,braithwaite_axisymmetric_2009,frieben_equilibrium_2012}} - meaning that mixed configurations like the twisted-torus are more likely \citep{ciolfi_2013,uryu_equilibrium_2014,pili_axisymmetric_2014} - the stabilising role of a rigid crust and of the external magnetosphere is yet to be evaluated. Luckily these magnetic fields might have potentially observable consequences, which might offer us a way to constrain them: they have the ability to modify the torsional oscillations of NSs {\citep{samuelsson_2007,sotani_2015}}, alter the cooling properties of their crust {\citep{page_2004,aguilera_2008},} and act as a potential source of deformation {\citep{haskell_2008,gomes_2019}}.
\\\\
In addition to the uncertainties regarding the magnetic properties of NS interiors, also their internal composition, encoded by the equation of state (EoS), remains mostly unconstrained. In this sense, the observation of NSs of a mass higher than 2M$_\odot$,  the most massive NS observed to date, with a mass of $\sim$2.28M$_\odot$ \citep{kandel_2020} or a 2.08M$_\odot$ NS \citep{Fonseca_2021}, rejected the validity of many proposed EoS; furthermore the first observation of gravitational waves (GWs) emitted by a binary NS merger \citep{abbott_gw170817:_2017} set additional limits on their stiffness \citep{abbott_2018_1,Bauswein_2019}. Moreover, if the low-mass component of the recent binary coalescence event GW190814 \citep{abbott_2020} is interpreted as a NS, it would set extremely stringent limits on the maximum mass that a valid EoS must be able to reach \citep{kanakis_2021,godzieba_2021,Lim_Bhattacharya_Holt_Pati_2021,rather_2021,bombaci_2021}. {On the other hand, the recent results of the NICER telescope \citep{pang_2021,zhang_2021} allowed  tighter limits to be set on the possible EoS of NS by constraining their radius.}
This uncertainty is further enhanced by the fact that the strong magnetic fields inside NSs directly affect their particle composition, for example, by determining the presence of exotic particles; thus, their interplay may have a key role in answering particle physics questions such as the hyperon puzzle \citep{zdunik_2013,chatterjee_2015}, the Delta puzzle \citep{cai_2015,Drago_Lavagno_Pagliara_Pigato_2016}, the hadron-quark phase transition \citep{avancini_2012,ferreira_2014,costa_2014,roark_2018,lugones_2019}, and the possibility of the existence of a superconducting phase \citep{ruderman_1995,lander_2013,haskell_2018}.
\\\\
Understanding and having the ability to constrain the interplay of a strong magnetic field with the EoS in determining the structure and properties of NSs is thus of great importance to advance our knowledge of these objects. Given that any time-varying  deformation leads to the emission of continuous GWs (CGWs), gravitational astronomy can offer an independent constraint also on the interior magnetic field of a NS {\citep{bonazzola_1996,yoshida_2002,lapiedra_2003,haskell_2008,gomes_2019,sieniawska_2019,abbott_ellipticity_2020,dergachev_2020,frederick_2021,cieslar_2021}}.
Among the many physical processes which can induce a deformation on the shape of a NS, we recall that mountains can form on their surface due to crustal deformations \citep{haskell_2006,usho_2000} or due to magnetic burial in accretion processes  \citep{melatos_2005}; also, oscillation modes such as the r-mode can develop an instability \citep{andersson_1998}, leading to the emission of CGWs \citep{abbott_rmodes_2021}; finally, the magnetic field itself can be the source of a global quadrupole \citep{bocquet_rotating_1995,cutler_2002,oron_relativistic_2002,Dall'Osso_Shore+09a,frieben_equilibrium_2012,pili_axisymmetric_2014,gomes_2019}. In the latter case, all that is needed to produce CGWs is a NS whose magnetic axis is not aligned to the rotation axis. It was shown \citep{cutler_2002} that a strong toroidal magnetic field would force the NS to develop an instability, flipping it to an orthogonal rotator, thus maximising its emission in terms of CGWs. Indeed \citet{lander_2018} showed that for parameters close to those of the observed population of NSs, most configurations are expected to evolve to orthogonal rotators. Moreover, it was found that
accretion can lead to an increase in the misalignment between the rotation
axis and the magnetic axis \citep{biryukov_2021}.  Since magnetic fields are supposed to be stronger in the deep interior, the magnetic deformation of a NS offers a way to probe the conditions in their core, as opposed to other potential sources of the deformation (including rotation and tidal forces), which mostly act on their outer layers. Unfortunately, only magnetic fields of strengths of $B \gtrsim 10^{14}$G, which is much higher than the surface magnetic field observed in regular pulsars \citep{atnf_2005}, can cause a significant deviation from spherical symmetry \citep{haskell_2008}. Barring the possibility of an interior superconducting phase, which could substantially enhance the effectiveness of the magnetic field in deforming the NS \citep{cutler_2002,akgun_2008,lander_2012,lander_2014}, similar fields are likely to be found only inside magnetars and newly born proto-NSs. Given the long spin period of observed classical magnetars, this leaves newly born proto-NSs and millisecond magnetars \citep{dallosso_2021} as the most promising sources of significant CGWs emission. However, as we show, close millisecond pulsars endowed with a superconducting core may be within the reach of third-generation GW detectors.
\\\\
Given the compactness of a NS (their typical radii are just about three times  larger that the Schwarzschild radius of a black hole of the same mass), any meaningful estimate of their role as potential CGWs sources requires them to be modeled in the strong gravitational field regime. With regard to this point, even if general relativity (GR) remains today the best theory to describe gravitation in the strong field regime [if we neglect the long-standing debate about quantum gravity \citep{bars_is_1989,deser_infinities_2000}], it has been established that our understanding, within its framework, of the gravitational interaction on galactic and cosmological scales presents some issues \citep{papantonopoulos_modifications_2015}. While most of the research has focussed on the introduction of a dark sector to account for some of these problems \citep{trimble_existence_1987,peebles_cosmological_2003}, the other possibility is to consider modifications of GR, leading to the introduction of alternative theories of gravity \citep{capozziello_extended_2011}.
\\\\
Among the infinitely many  possible alternatives to GR, scalar-tensor theories (STTs) \citep{brans_machs_1961} have attracted a lot of interest for their simplicity, as they are the most simple alternatives to GR; for more details, see Sect.~1.2 of \citet{papantonopoulos_modifications_2015}. This is  because they are predicted to be the low-energy limit of some possible theories of quantum gravity \citep{damour_runaway_2002} since most of them respect the weak equivalence principle \citep{will_confrontation_2014}, which has been extremely well tested \citep{touboul_microscope_2017} and  they seem to be free of pathologies affecting other GR extensions \citep{defelice_2006,defelice_2010,bertolami_2016}. Interestingly, some STTs show a non-perturbative strong field effect, named spontaneous scalarisation \citep{damour_nonperturbative_1993}, which, while not constrained by observations in the weak-field limit \citep{shao_constraining_2017,voisin_2020}, leads to strong deviations from GR in the physics of material compact objects. Among the enrichments in NS phenomenology, we name the emission of additional modes of GWs {\citep{eardley_1973,chatziioannou_2012,pang_2020}}; a modified relation between the NS mass, radius, and central density; different binary NS (BNS) merger dynamics \citep{shibata_coalescence_2014}; a variation in the frequency of NS normal modes \citep{sotani_2005}, in their tidal and rotational deformation \citep{pani_2014,doneva_iq_2014}, and  in their magnetic deformation \citep{soldateschi_2021}; a modification of light propagation in their surroundings \citep{bucciantini_2020}. Even if spontaneous scalarisation in the simplest versions of STTs (i.e. massless ones) have been almost completely ruled out thanks to measurements in binary pulsar systems \citep{will_confrontation_2014,shao_constraining_2017,voisin_2020}, this phenomenon remains worth investigating: firstly because it cannot be ruled out for massive fields or fields with screening potentials \citep{ramazanoglu_spontaneous_2016,yaza_2016,doneva_rapidly_2016,staykov_2018,xu_2020,rosca_2020}; secondly, because it is the prototype of  other kinds of scalarisation \citep{salgado_1998,barausse_neutron-star_2013,ramazanoglu_spontaneous_2016,rama_2017,silva_2018,andreou_2019}; thirdly, because multi-dimensional and time dependent numerical algorithms developed in GR can easily be extended to handle STTs \citep{novak_spherical_1998,shibata_coalescence_2014,gerosa_numerical_2016,soldateschi_2020}, which cannot be said of virtually all other GR alternatives.
Unfortunately, part of the phenomenology of STTs is degenerate with  the EoS of NSs, for example, regarding their mass-radius relation or deformabilities. For this reason, it is important to find ways to disentangle them, such that a more unambiguous interpretation of observations could be made possible.
\\\\
In this study, we build upon the works of \citet{soldateschi_2020} and \citet{soldateschi_2021} (hereafter \citetalias{soldateschi_2020} and \citetalias{soldateschi_2021}, respectively), presenting a comprehensive study of the magnetic deformability of NSs, both for poloidal and toroidal magnetic field configurations, for a large sample of different EoS, in GR and STTs, for all masses in the stable range above 1M$_\odot$. In this regard, our work aims to build upon and complement several previous studies: \citet{frieben_equilibrium_2012}, who investigated the magnetic deformation of toroidal configurations in GR for various EoS at a single NS mass of 1.4M$_\odot$; \citet{pili_axisymmetric_2014}, who investigated toroidal, poloidal, and mixed configurations  in GR  for various masses  but only for a simple polytropic EoS; \citetalias{soldateschi_2021}, where we studied, for the first time, the problem of the magnetic deformation of NSs endowed by spontaneous scalarisation with a massless scalar field for a simple polytropic EoS, for toroidal and poloidal magnetic fields and for various masses in the stable range.

Our aim is twofold: firstly, it is to better understand the interplay between different EoS and the magnetic field of a NS and explore whether some kind of EoS-independent relation between the NS deformation and its observable quantities exists, such as mass and radius. This would help to shed some light onto the properties of the internal magnetic field of NSs or set limits on their possible CGW emission. Secondly, we  look for similar relations in the case of a scalarised NS, in the case of which EoS-independent scalings could be useful for disentangling the effect of the scalar field to that of the EoS. Let us recall here that despite their over-simplicity (see the discussion in \citetalias{soldateschi_2021}, also concerning limitations related to the equilibrium formalism), purely toroidal and purely poloidal cases represent the  two extrema of the much larger space of possible magnetic configurations {\citep{akgun_2013,mastrano_2013,mastrano_2015}} and that for any given magnetic field energy, they maximize the NS quadrupolar deformation, be it prolate or oblate (as in mixed configurations), the two components are expected to balance each other out; see {\citet{tayler_1980}}. Thus, estimates in these two limits form reliable bounds on possible CGWs emission scenarios.
\\\\
This paper is structured as follows. In Sect.~\ref{sec:grstt}, we make a brief recap of the formalism of magnetohydrodynamical (MHD) equilibria in GR and in STTs. In Sect.~\ref{sec:eos}, we introduce the EoS that we considered in this work, along with the rationale behind our choice. In Sect.~\ref{sec:modsetup}, we define the magnetic deformation and explain the setup of our numerical code and of the physical parameters we adopted. In Sect.~\ref{sec:results}, we show our results for the magnetic deformation of NSs in GR and STTs; first for `traditional' NSs and then for strange quark stars. In Sect.~\ref{sec:constr}, we elaborate on the possible applications of our work for constraining the NS EoS and magnetic structure and their detectability by way of CGWs. Finally, we present our conclusions in Sect.~\ref{sec:conclusions}.

\section{Neutron stars in general relativity and beyond}\label{sec:grstt}
In the following, we assume a signature $\{ -,+,+,+ \}$ for the spacetime metric and use Greek letters $\mu, \nu, \lambda, \dots$ (running from 0 to 3) for 4D spacetime tensor components, while Latin letters $i, j, k, \dots$ (running from 1 to 3) are employed for 3D spatial tensor components. We use dimensionless units, where $c=G=\mathrm{M}_\odot = 1$, and we absorb the $\sqrt{4\pi}$ factors in the definitions of the electromagnetic quantities. All quantities $f$ calculated in the Einstein frame (E-frame) are denoted with a bar ($\bar{f}$), while all quantities calculated in the Jordan frame (J-frame) are denoted with a tilde ($\tilde{f}$). {For more information, see \citet{sotiriou_2008} for the definition of the J-frame and the E-frame}. Moreover, we focus on the case of static, axisymmetric configurations; see \citet{pili_axisymmetric_2014,pili_general_2015,pili_general_2017} for a more thorough derivation of the equations dictating the structure of magnetised, axisymmetric models, both in the static and in the stationary rotating case, in GR; see \citetalias{soldateschi_2020} for the derivation, in the static case, in STTs.

\subsection{General relativity}\label{subsec:gr}
In the case of static, axisymmetric configurations, within a good level of accuracy \citep{oron_relativistic_2002,shibata_magnetohydrodynamics_2005,dimmelmeier_non-linear_2006,ott_rotating_2007,bucciantini_general_2011,pili_axisymmetric_2014,pili_general_2017}, the spacetime metric can be approximated using the conformally flat condition (CFC) \citep{wilson_mathews_2003,isenberg_waveless_2008}. Using spherical-like coordinates $x^\mu = [t,r,\theta,\phi]$, the line element is
\begin{equation}\label{eq:3+1metric}
    g_{\mu \nu}dx^\mu dx^\nu = -\alpha^2dt^2 + \psi^4\left[ dr^2 + r^2d\theta^2 + r^2\sin^2\theta d\phi^2 \right] \; ,
\end{equation}
where $g_{\mu \nu}$ is the spacetime metric, having determinant $g$, $\alpha (r,\theta)$ is called the lapse function and $\psi (r,\theta)$ is called the conformal factor. The energy-momentum tensor for a magnetised ideal fluid is \citep{bucciantini_fully_2013,chatterjee_2015,franzon_internal_2016,del_zanna_fast_2016,del_zanna_chiral_2018,tomei2020}
\begin{equation}\label{eq:tmunu}
    T^{\mu \nu}_\mathrm{p} = \left( \rho + \varepsilon + p \right) u^\mu u^\nu + p g^{\mu \nu} + F^\mu _\lambda F^{\nu \lambda} - \frac{1}{4}F^{\lambda \kappa} F_{\lambda \kappa} g^{\mu \nu} \; ,
\end{equation}
where $\rho$ is the rest mass density, $\varepsilon$ is the internal energy density, $p$ is the pressure, $u^\mu$ is the four-velocity, and $F^{\mu \nu}$ is the Faraday tensor. The subscript `p' stands for the `physical' - fluid and electromagnetic - fields, to distinguish them from the scalar field which is introduced in Sect.~\ref{subsec:stt}.
Einstein's equations can be recast using the 3+1 formalism \citep{alcubierre_introduction_2008,gourgoulhon_3+1_2012} and under our assumptions, they take the form of two Poisson-like equations: one for $\psi$ and one for $\alpha$ \citep{pili_axisymmetric_2014}.
As for the MHD quantities, introducing the pseudo-enthalpy, $h$, related to the pressure and density by ${\rm d} \ln{h} = {\rm d} p /(\rho+\varepsilon+p)$, Euler's equation takes the form of the `generalised Bernoulli integral' \citep{pili_axisymmetric_2014}:
\begin{equation}\label{eq:polobernoulli}
        \ln \left( \frac{h}{h_{\mathrm c}} \right) + \ln \left( \frac{ \alpha}{ \alpha _{\mathrm c}} \right) - \mathcal{M}=0 \; ,
\end{equation}
where $\mathcal{M}$ is the magnetisation function and $h_{\mathrm c}$ and $\alpha _{\mathrm c}$ are the values of $h$ and $\alpha$ at the centre of the star, respectively (we have assumed $\mathcal{M}_{\mathrm c}=0$). Among the various possible choices \citep{pili_axisymmetric_2014,bucciantini_role_2015}, following \citetalias{soldateschi_2021}, in this work, the magnetisation function in the case of a purely poloidal magnetic field is taken to be:
\begin{equation}\label{eq:polmag}
        \mathcal{M}=k_{\mathrm{pol}}A_\phi \; ,
\end{equation}
where $k_{\mathrm{pol}}$ is called the poloidal magnetisation constant and $A_\phi$ is the $\phi$-component of the vector potential. For a purely toroidal magnetic field, we instead use:
\begin{equation}\label{eq:tormag}
    \mathcal{M} = - k_\mathrm{tor}^2 (\rho +\varepsilon + p) \mathcal{R}^2 \; ,
\end{equation}
where $k_{\mathrm{tor}}$ is the toroidal magnetisation constant and $\mathcal{R}^2~=~\alpha ^2 \psi ^4 r^2 \sin ^2 \theta$, corresponding to a toroidal magnesation index equal to unity. In the former case, the poloidal components of the magnetic field are found by solving the Grad-Shafranov equation \citep{del_zanna_exact_1996,pili_general_2017} for $A_\phi$; in the latter, the toroidal component is simply directly proportional to $(\rho +\varepsilon + p) \mathcal{R}^2/\alpha$. The limits and advantages of this approach based on the `generalised Bernoulli integral'  are amply discussed in \citetalias{soldateschi_2021}.
Finally, an EoS is needed to close the system of equations, relating the fluid quantities of $\rho, p, \varepsilon$. We describe our selection of EoS in Sect.~\ref{sec:eos}.
\subsection{Scalar-tensor theories}\label{subsec:stt}
In STTs, according to the `Bergmann-Wagoner formulation', the action in the J-frame is \citep{bergmann_comments_1968,wagoner_scalar-tensor_1970,berti_testing_2015}:
\begin{equation}\label{eq:joract}
\begin{split}
        \tilde{S}&= \frac{1}{16\pi}\int \mathrm{d}^4x \sqrt{-\tilde{g}}\left[ \varphi \tilde{R} - \frac{\omega (\varphi)}{\varphi} \tilde{\nabla} _\mu \varphi \tilde{\nabla} ^\mu \varphi - U(\varphi) \right] +\\
        &+ \tilde{S}_\mathrm{p}\left[ \tilde{\Psi} , \tilde{g}_{\mu \nu} \right]  \; ,
\end{split}
\end{equation}
where $\tilde{g}$ is the determinant of the spacetime metric $\tilde{g}_{\mu \nu}$, $\tilde{\nabla} _\mu$ its associated covariant derivative, $\tilde{R}$ its Ricci scalar, while $\omega (\varphi)$ and $U(\varphi)$ are, respectively, the coupling function and the potential of the scalar field $\varphi$, and $\tilde{S}_\mathrm{p}$ is the action of the physical fields $\tilde{\Psi}$. We focus on massless scalar fields, thus $U(\varphi)=0$. The action in the E-frame is obtained through the conformal transformation $\bar{g}_{\mu \nu}~=~\mathcal{A}^{-2}(\chi) \tilde{g}_{\mu \nu}$, where $\mathcal{A}^{-2}(\chi) =\varphi (\chi)$ and $\chi$ are redefinitions of the coupling function and of the scalar field in the E-frame, respectively. The two scalar field definitions are related by:
\begin{equation}
\frac{{\mathrm d}\chi}{{\mathrm d}\ln \varphi} = \sqrt{\frac{\omega{(\varphi)} +3}{4}} \; .
\end{equation}
The equation governing the scalar field in the E-frame is \citepalias{soldateschi_2020}:
\begin{equation}\label{eq:scal}
        \Delta \chi = -4\pi \bar{\psi}^4 \alpha _\mathrm{s}(\chi) \mathcal{A}^4 \tilde{T}_\mathrm{p}-\partial \ln \left( \bar{\alpha} \bar{\psi} ^2 \right) \partial \chi \; ,
\end{equation}
where $\Delta = f^{ij} \hat{\nabla} _i \hat{\nabla} _j$ and $\hat{\nabla} _i$ are, respectively, the 3D Laplacian and nabla operator of the flat space metric $f_{ij}$, $\partial f \partial g~=~\partial _r f \partial _r g + (\partial _\theta f \partial _\theta g)/r^2$, $\alpha _\mathrm{s}(\chi) = {\mathrm d} \ln \mathcal{A}/{\mathrm d}\chi,$ and $\tilde{T}_\mathrm{p} = 3\tilde{p}- \tilde{\varepsilon}-\tilde{\rho}$ is the trace of the J-frame energy momentum tensor of the physical fields.

We adopted an exponential coupling function $\mathcal{A}(\chi)~=~\exp \{ \alpha _0 \chi + \beta _0 \chi ^2 /2 \} $, first introduced in \citet{damour_nonperturbative_1993}. The $\alpha _0$ parameter regulates the weak-field effects, while the $\beta _0$ parameter controls spontaneous scalarisation. The tightest observational constraints to date require that, for massless scalar fields, $|\alpha _0| \lesssim 1.3\times 10^{-3}$ and $|\beta _0| \gtrsim 4.3$ \citep{voisin_2020}; for massive scalar fields, lower values of $\beta _0$ are still allowed \citep{doneva_rapidly_2016}, as long as the screening radius is smaller than the binary separation. We emphasise here that results found in a massless STT for the internal structure of a NS are also valid for screened STTs as long as the screening radius is larger than the NS radius.

While the scalar field is non-minimally coupled to the spacetime metric in the J-frame, it is minimally coupled in the E-frame. However, the weak equivalence principle holds only in the J-frame, and not in the E-frame {\citep{sotiriou_2008}}. For these reasons, Einstein's equations retain their usual form in the E-frame, while MHD equations have the usual expression in the J-frame. Following this reasoning, all the equations described in Sect.~\ref{subsec:gr} hold also in STTs, with the caveat of expressing them in the appropriate frame and adding the energy-momentum tensor of the scalar field to the physical one in the Einstein's equations in the E-frame. This translates in adding the source terms of the scalar field to the sources of the equations for $\bar{\psi}$ and $\bar{\alpha}$. In the E-frame, the energy-momentum tensor of the scalar field is expressed as
\begin{equation}\label{eq:tmunuscal}
        \bar{T}_{\mathrm{s}}^{\mu \nu}=\frac{1}{4\pi}\left[ \bar{\nabla} ^\mu \chi \bar{\nabla} ^\nu \chi - \frac{1}{2} \bar{g}^{\mu \nu} \bar{\nabla} _\lambda \chi \bar{\nabla} ^\lambda \chi \right]\; .
\end{equation}
Given the ambiguity of the two frames in STTs, a clarification is due.
In the following, when referring to the STT version of fluid and electromagnetic quantities (as with the magnetic field, $B$, the magnetic energy, $\mathcal{H}$, the density, $\rho$, but also the circumferential radius, $R_\mathrm{c}$), we always refer to their J-frame value; instead, when referring to metric and scalar field quantities (like the Komar mass, $M_\mathrm{k}$, the scalar charge, $Q_\mathrm{s}$ and the quadrupolar deformation, $e$), we always refer to their E-frame value. For this reason, and for legibility, henceforth we drop the bar and the tilde notation. Whenever necessary for reasons of ambiguity, we do restore the bar and the tilde notation to specify the frame in which quantities are computed.

\section{Selection of equations of state}\label{sec:eos}
We chose a selection of 13 different EoS that span a diverse range of calculation methods and particle contents: from zero-temperature and $\beta$-equilibrium purely nucleonic EoS to more particle-rich and finite temperature ones, considering also EoS for strange quark stars and a polytropic one. Moreover, all the EoS we used, except the polytropic one (which was used as a comparison to previous works in the literature), were chosen based on the fact that they were still in line with the observational constraints: 1) reaching a maximum mass of at least $\sim$2.05M$_\odot$; 2) satisfying various nuclear physics constraints \citep{Fortin_Providencia_Raduta_Gulminelli_Zdunik_Haensel_Bejger_2016};  3) not too stiff \citep{chaves_2019}; and 4) providing a radius between $\sim$10km and $\sim$14km for 1.4M$_\odot$ mass models {\citep{bauswein_2017,Bombaci_Logoteta_2018,Kim_Kim_Sung_Lee_Kwak_2021,miller_2021,raaijmakers_2021,riley_2021}}. The scope of this work is not focussed on analysing a very large sample of EoS, but a sample as diverse as possible in terms of physics, particle content, and computational methods, within a range that is in reasonable agreement with present constraints. In this sense, we note that the existence of strange quark stars is considered a possible way out of the hyperon and Delta puzzles, according to which the high densities reached at the centre of hadronic stars would soften the EoS, causing a less-than-2M$_\odot$ maximum mass to be reached \citep{Drago_Lavagno_Pagliara_Pigato_2016}; such stars could be observed, for example, through the emission of multimessenger signals caused by the conversion of a NS to a strange quark star \citep{Kuzur_Mallick_R_Singh_2021}.
\\\\
In the following, we summarise the main feature of each EoS, grouped according to the particle content: `nucleonic' for EoS that contain only $npe\mu$ particles; `hyperonic' for EoS that contain also hyperons; `quarkionic' for EoS that contain an $uds$ quark matter domain treated with the Nambu-Jona-Lasinio model; `strange quark matter' for EoS containing $uds$ quarks treated with the MIT bag model or perturbative QCD. We note that this last class of EoS predicts that for certain values of their parameters, the quark matter phase has an energy-per-baryon at zero pressure lower than that of $^{56}$Fe \citep{Bodmer_1971,Witten_1984}, leading to the possible existence of strange quark stars \citep{Madsen_1991,Glendenning_2000}.
For convenience, we provide  a detailed description of the selected EoS here, characterising their main physical properties and assumptions.
\subsection{Nucleonic }
    APR: Zero temperature and $\beta$-equilibrium $npe\mu$ matter by \citet{akmal_1998}, computed using variational techniques with the two-nucleon interaction A18, the boost correction due to nearby nucleons $\delta v$ and the three-body interaction term UIX* in the baryon number density range of $7.6\times 10^{-2}~<~n_\mathrm{b}/\mathrm{fm}^{-3}~<~1.34$ (the liquid core). We note that this EoS is usually called APR4 in the literature. The inner crust, in the range of $2.1\times 10^{-4}~<~n_\mathrm{b}/\mathrm{fm}^{-3}~<~7.6\times 10^{-2}$, is calculated with the SLy4 EoS \citep{Douchin_Haensel_2001} and attached to the liquid core; the outer crust, in the range of $8\times 10^{-15}~<~n_\mathrm{b}/\mathrm{fm}^{-3}~<~2.1\times 10^{-4}$, is from \citet{Baym_Pethick_Sutherland_1971} and attached to the inner crust. Data is taken from the CompOSE database \citep{Typel_Oertel_Klaehn_2013}\footnote{See the CompOSE databse website for details: \url{https://compose.obspm.fr/home}.}\footnote{We note that the pseudo-enthalpy $h$ is not available in the CompOSE database, and was thus computed when needed.}.
\\\\
    SLY9: Zero temperature and beta-equilibrium $npe\mu$ matter unified EoS by \citet{Gulminelli_Raduta_2015}, taken from the CompOSE database where it is named `RG(SLY9)'. Cluster energy functionals are those of \citet{Danielewicz_Lee_2009}. The high density part $10^{-7}~<~n_\mathrm{b}/\mathrm{fm}^{-3}~<~1.51$ is calculated using the effective interaction SLy9 \citep{Chabanat_1995}. This is the only Skyrme-type EoS that seems to satisfy various nuclear physics constraints \citep{Fortin_Providencia_Raduta_Gulminelli_Zdunik_Haensel_Bejger_2016}. The low density part $8.6\times 10^{-11}~<~n_\mathrm{b}/\mathrm{fm}^{-3}~<~10^{-7}$ has been added to the original CompOSE table using the EoS by \citet{Togashi_Nakazato_Takehara_Yamamuro_Suzuki_Takano_2017}. Beware that the EoS from \citet{Fortin_Providencia_Raduta_Gulminelli_Zdunik_Haensel_Bejger_2016} (thus, also from CompOSE) has been joined at $n_\mathrm{b}=0.04$fm$^{-3}$ in a thermodinamically unstable way (Raduta private communication), because the pressure drops with increasing density. We adjusted this point, but it still results in a minor density jump in the models.
\\\\
    BL2: Zero temperature and beta–equilibrium $npe\mu$ matter obtained using realistic two-body and three-body nuclear interactions derived in the framework of chiral perturbation theory, including the $\Delta (1232)$ isobar intermediate state. The high density part $8\times 10^{-2}~<~n_\mathrm{b}/\mathrm{fm}^{-3}~<~1.29$ is from \citet{Bombaci_Logoteta_2018}, where it is named N3LO$\Delta$+N2LO$\Delta$, and is derived using Brueckner-Bethe-Goldstone quantum many-body theory in the Brueckner-Hartree-Fock approximation. The crust EoS is taken from \citet{Douchin_Haensel_2001}, in the density range of $7.9\times 10^{-15}~<~n_\mathrm{b}/\mathrm{fm}^{-3}~<~8.0\times 10^{-2}$. Data is taken from the CompOSE database, where it is named `BL\_EOS with crust'.
\\\\
    DDME2: Zero temperature and beta-equilibrium EoS by \citet{Fortin_Providencia_Raduta_Gulminelli_Zdunik_Haensel_Bejger_2016}, computed in the high density range of $2.5\times 10^{-4}~<~n_\mathrm{b}/\mathrm{fm}^{-3}~<~1.2$ with a relativistic-mean-field theory model where nucleons interact via the exchange of $\sigma, \omega, \rho$ mesons with density dependent meson-nucleon couplings by \citet{lalazissis_2005}, where it is named `DD-ME2'. The outer crust follows the SLy9 EoS by \citet{Chabanat_1995} in the high density regime $10^{-7}~<~n_\mathrm{b}/\mathrm{fm}^{-3}~<~2.5\times 10^{-4}$, while it is taken from \citet{Douchin_Haensel_2001} for the low desity regime $n_\mathrm{b}~<~10^{-7}$fm$^{-3}$. Data tables are found in the supplemental materials of \citet{Fortin_Providencia_Raduta_Gulminelli_Zdunik_Haensel_Bejger_2016}.
\\\\
    NL3$\omega \rho$: Zero temperature and beta-equilibrium EoS by \citet{Fortin_Providencia_Raduta_Gulminelli_Zdunik_Haensel_Bejger_2016}, computed in the high density range of $3\times 10^{-4}~<~n_\mathrm{b}/\mathrm{fm}^{-3}~<~1.2$ with a relativistic-mean-field theory Walecka model where nucleons interact via the exchange of $\sigma, \omega, \rho$ mesons with non-linear meson-meson coupling terms by \citet{Horowitz_Piekarewicz_2001}. The outer crust follows the SLy9 EoS by \citet{Chabanat_1995} in the high-density regime $10^{-7}~<~n_\mathrm{b}/\mathrm{fm}^{-3}~<~3\times 10^{-4}$, while it is taken from \citet{Douchin_Haensel_2001} for the low desity regime $n_\mathrm{b}~<~10^{-7}$. Data tables are found in the supplemental materials of \citet{Fortin_Providencia_Raduta_Gulminelli_Zdunik_Haensel_Bejger_2016}.
\\\\
    SFH: finite $0.1$MeV temperature and beta–equilibrium $npe$ matter EoS from \citet{Steiner_Hempel_Fischer_2013} in the range of $10^{-12}~<~n_\mathrm{b}/\mathrm{fm}^{-3}~<~1.9$. It is calculated with the statistical model with excluded volume and interactions of \citet{Hempel_Schaffner-Bielich_2010}, with relativistic mean-field-theory interactions SFHo. A power-law extrapolation at low densities is used to avoid the effects of finite temperature, which emerge as a pressure saturation around $n_\mathrm{b}\approx 10^{-10}$. Data is taken from the CompOSE database\footnote{See also the website \url{https://astro.physik.unibas.ch/en/people/matthias-hempel/equations-of-state/} for more details.}, where it is named `SFHO (with electrons)'.
\subsection{Hyperonic}
    DDME2-Y: Equivalent to the DDME2 EoS, but with the inclusion of the six lightest hyperons $\Lambda ^0$ (for $n_\mathrm{b}>0.34$fm$^{-3}$), $\Sigma ^{0,\pm}$ (for $n_\mathrm{b}>0.41$fm$^{-3}$) and $\Xi ^\pm$ (for $n_\mathrm{b}>0.37$fm$^{-3}$) with the hidden strangeness vector-isoscalar $\phi$ meson. Hyperon-meson coupling coefficients are calculated according to \citet{Fortin_Providencia_Raduta_Gulminelli_Zdunik_Haensel_Bejger_2016} using relativistic-mean-field theory calculations. Data tables are found in the supplemental materials of \citet{Fortin_Providencia_Raduta_Gulminelli_Zdunik_Haensel_Bejger_2016}.
\\\\
    NL3$\omega \rho$-Y: equivalent to the NL3$\omega \rho$ EoS, but with the inclusion of the six lightest hyperons $\Lambda ^0$ (for $n_\mathrm{b}>0.31$fm$^{-3}$), $\Sigma ^{0,\pm}$ (for $n_\mathrm{b}>0.49$fm$^{-3}$) and $\Xi ^\pm$ (for $n_\mathrm{b}>0.34$fm$^{-3}$) with the hidden strangeness vector-isoscalar $\phi$ meson. Hyperon-meson coupling coefficients are calculated according to \citet{Fortin_Providencia_Raduta_Gulminelli_Zdunik_Haensel_Bejger_2016} using relativistic-mean-field theory calculations. Data tables are found in the supplemental materials of \citet{Fortin_Providencia_Raduta_Gulminelli_Zdunik_Haensel_Bejger_2016}.
\subsection{Quarkionic}
    BH8: Zero-temperature and $\beta$-equilibrium unified EoS by \citet{baym_2018}\footnote{See also the website \url{https://user.numazu-ct.ac.jp/~sumi/eos/index.html\#QHC18} for more details.}. The EoS is divided into four distinct domains: the crust ($1.6\times 10^{-10}~<~n_\mathrm{b}/\mathrm{fm}^{-3}~<~4.16\times 10^{-2}$) is taken from \citet{Togashi_Nakazato_Takehara_Yamamuro_Suzuki_Takano_2017}; the nuclear liquid ($4.16\times 10^{-2}~<~n_\mathrm{b}/\mathrm{fm}^{-3}~<~0.32$) is taken from \citet{akmal_1998}; the hadron-quark crossover and quark matter domains ($0.32~<~n_\mathrm{b}/\mathrm{fm}^{-3}~<~1.6$) are taken from \citet{baym_2018}. It is necessary to note that the nuclear liquid contains a pion condensate, where $n_\mathrm{b}$ jumps from 0.21fm$^{-3}$ to 0.245fm$^{-3}$. The quark matter EoS (including up, down, and strange quarks) is calculated using the Nambu-Jona-Lasinio model within the mean field approximation. Its parameters $(g_\mathrm{V},H)$, which quantify the strength of the repulsive density-density interaction and the attractive pairing interaction between quarks respectively, have been chosen to be $(0.80,1.50)G_\mathrm{s}$, where $G_\mathrm{s}$ is the scalar coupling of the Nambu-Jona-Lasinio model for quark matter, which are compatible with hadron physics. Data is taken from the CompOSE database, where it is named `QHC18'.
\\\\
    BF9: zero-temperature and $\beta$-equilibrium unified EoS by \citet{Baym_Furusawa_Hatsuda_Kojo_Togashi_2019}. The EoS is divided into three distinct domains: the crust and nuclear liquid ($7.58\times 10^{-11}~<~n_\mathrm{b}/\mathrm{fm}^{-3}~<~0.32$) are taken from \citet{Togashi_Nakazato_Takehara_Yamamuro_Suzuki_Takano_2017}, while the hadron-quark crossover and quark matter domains ($0.32~<~n_\mathrm{b}/\mathrm{fm}^{-3}~<~1.58$) are taken from \citet{Baym_Furusawa_Hatsuda_Kojo_Togashi_2019}, model-B (corresponding to $g_\mathrm{V}=0.80G_\mathrm{s}$ and $H=1.49G_\mathrm{s}$). There are no phase transitions. The quark matter EoS (including up, down, and strange quarks) is calculated using the Nambu-Jona-Lasinio model within the mean field approximation. Data is taken from the CompOSE database, where it is named `QHC19-B'.
\subsection{Strange quark matter}
\begin{figure}
        \centering
         \includegraphics[width=\columnwidth]{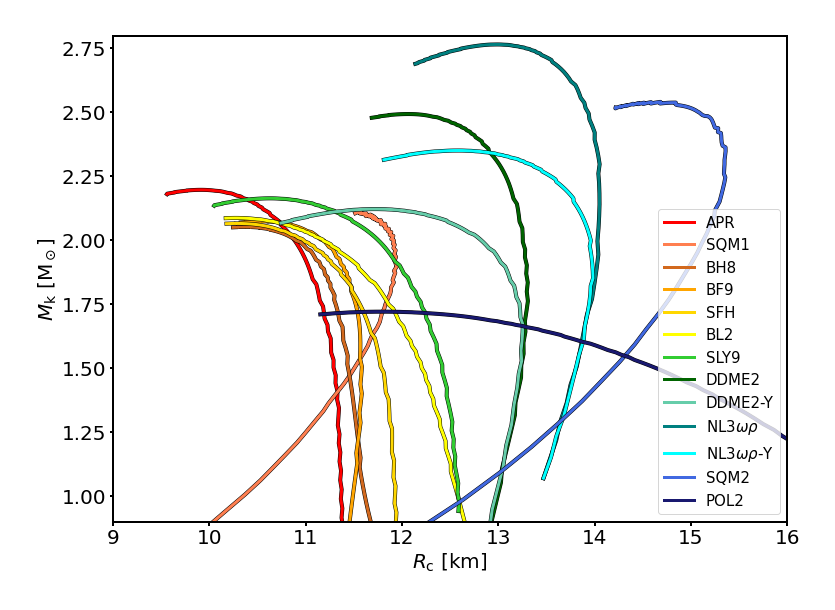}

         \includegraphics[width=\columnwidth]{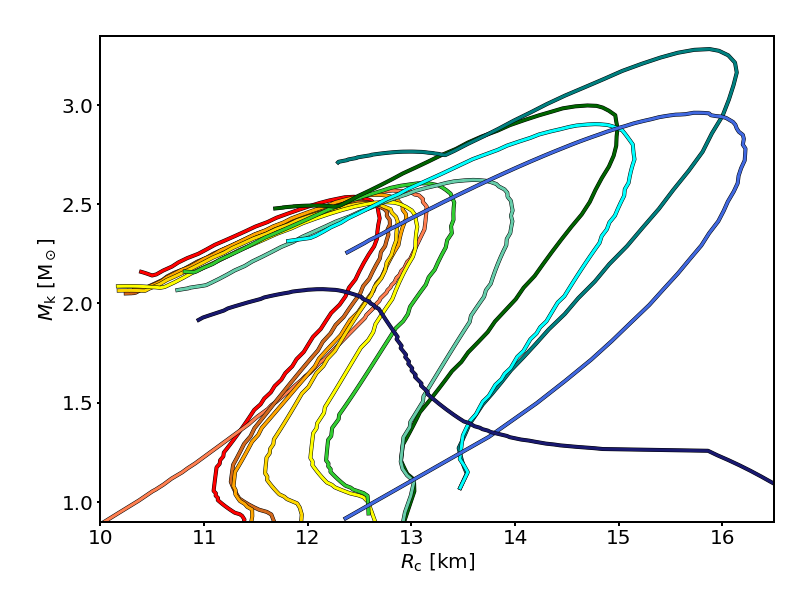}
         \caption{Komar mass $M_\mathrm{k}$ against circumferential radius $R_\mathrm{c}$ for un-magnetised, static models of NSs computed with the EoS described in Sect.~\ref{sec:eos} in GR (top plot) and in STTs with $\beta_0=-6$ (bottom plot). The EoS are colour-coded, and ordered in the legend, according to the compactness $C=M_\mathrm{k}/R_\mathrm{c}$ calculated at $M_\mathrm{k}=1.4$M$_\odot$ in GR: red for the highest compactness and blue for the lowest compactness.}
         \label{fig:mrrel}
 \end{figure}

    SQM1: obtained by \citet{Alcock_Farhi_Olinto_1986} [see also \citet{farhi_1984}] using the MIT bag model, setting the strong interaction coupling constant $\alpha_\mathrm{c}=0$ and assuming the strange quark mass $m_\mathrm{s}=0$, leading to a vanishing electron density. No pairing is present. The MIT bag constant was chosen to be $B=(141.4\mathrm{MeV})^4$, corresponding to a zero pressure density of $n_\mathrm{b}=0.259$fm$^{-3}$ and a maximum mass of $\sim 2.1$M$_\odot$.
\\\\
    SQM2: based on the parametrisation of the perturbative QCD calculations at finite chemical potential according to \citet{Fraga_Kurkela_Vuorinen_2014} [see also \citet{Drago_Lavagno_Pagliara_Pigato_2016,pili_quark_2016} for other works using this EoS], setting the strong interaction coupling constant at the $Z$ mass scale $\alpha _\mathrm{c}=0.118$ and assuming the strange quark mass $m_\mathrm{s}=94$MeV (at the 2GeV scale). The scale parameter $X$, which is the ratio between the renormalisation scale and the baryon chemical potential, have been chosen to be $X=3.5$, for which the maximum mass of quark stars is 2.54M$_\odot$. Quark matter is unpaired. The zero-pressure density is $n_\mathrm{b}=0.140$fm$^{-3}$.
\\\\
Finally, we also used the POL2 polytropic EoS, widely used in previous literature \citep{bocquet_rotating_1995,kiuchi_relativistic_2008,frieben_equilibrium_2012,pili_axisymmetric_2014}: $p=K_\mathrm{a} \rho ^{\gamma _\mathrm{a}}$, with $K_\mathrm{a} = 110$ (in dimensionless units) and $\gamma _\mathrm{a} = 2$. In the following nucleonic, hyperonic and quarkionic EoS  are  generally referred as `standard EoS', and NSs that are computed via these EoS are noted as `standard NS'.
\\\\
In Fig.~\ref{fig:mrrel}, we plot the Komar mass $M_\mathrm{k}$ against the circumferential radius $R_\mathrm{c}$ for models of un-magnetised, static NSs computed with the described EoS. The top panel refers to GR, while the bottom panel refers to a STT with $\beta_0=-6$ (see App.~C of \citetalias{soldateschi_2020} for the definitions of global quantities in the STT case). The maximum mass models, as well as the radius for the models having a Komar mass of 1.4M$_\odot$, are characterised for each EoS in Tab.~\ref{tab:eos}, in GR and in STT with $\beta_0=-6$. From Fig.~\ref{fig:mrrel} (top panel) we see that the NS radii have values ranging from $\sim$10km to $\sim$14km for most EoS, while the less compact SQM2, and especially the POL2, can reach radii higher than $15-16$km. The maximum masses are concentrated in the range of $\sim$2-2.2M$_\odot$ for most EoS, while they can reach the exceptionally high value of $\sim$2.77M$_\odot$ for the NL3$\omega \rho$ EoS, as well as the very low value of $\sim$1.72M$_\odot$ for the POL2 EoS. We note that the versions of the DDME2 and NL3$\omega \rho$ EoS containing hyperons, DDME2-Y and NL3$\omega \rho$-Y, respectively, detach from their non-hyperonic counterparts for models having central densities such that the six lightest hyperons may appear, corresponding to typical masses in the range of 1.70-1.75M$_\odot$. As is known \citep{Drago_Lavagno_Pagliara_Pigato_2016}, the appearance of hyperons causes a reduction of the maximum mass achievable by the NS.

\begin{table*}
\caption{Characterisation of the maximum mass models in GR (STT with $\beta _0=-6$), for each EoS described in Sect.~\ref{sec:eos} and plotted in Fig.~\ref{fig:mrrel}: $M^\mathrm{max}_\mathrm{k}$ is the Komar mass, $R^\mathrm{max}_{\mathrm c}$ is the circumferential radius, $\rho^\mathrm{max}_{\mathrm c}$ is the central density and $Q^\mathrm{max}_\mathrm{s}$ is the scalar charge. Moreover, for each EoS the circumferential radius for the model with $M_\mathrm{k}=1.4$M$_\odot$, $R_{1.4}$, is listed.}

\centering
\label{tab:eos}
\begin{tabular}{c c c c c c}
 \hline\hline
 \noalign{\smallskip}
  EoS & $M^\mathrm{max}_\mathrm{k}[\mathrm{M}_\odot]$ & $R^\mathrm{max}_{\mathrm c}[$km$]$ & $\rho^\mathrm{max}_{\mathrm c}[10^{15}$g cm$^{-3}]$ & $Q^\mathrm{max}_\mathrm{s}[$M$_\odot]$ & $R_{1.4}[$km$]$ \\ [0.5ex]
  \noalign{\smallskip}
 \hline
  \noalign{\smallskip}
  APR & 2.20(2.54) & 9.91(12.44) & 1.92(1.35) & (1.12) & 11.33(11.41) \\
  SQM1 & 2.11(2.57) & 11.57(12.82) & 1.55(2.21) & (1.27) & 11.33(11.46) \\
  BH8 & 2.05(2.51) & 10.36(12.52) & 1.88(1.50) & (1.15) & 11.48(11.54) \\
  BF9 & 2.07(2.55) & 10.54(12.69) & 1.81(1.48) & (1.18) & 11.58(11.61) \\
  SFH & 2.07(2.50) & 10.27(12.57) & 1.91(1.48) & (1.13) & 11.83(11.78) \\
  BL2 & 2.09(2.51) & 10.26(12.76) & 1.93(1.45) & (1.14) & 12.29(12.10) \\
  SLY9 & 2.16(2.61) & 10.63(13.15) & 1.78(1.35) & (1.19) & 12.47(12.28) \\
  DDME2 & 2.49(3.00) & 12.06(14.70) & 1.36(1.02) & (1.34) & 13.20(13.04) \\
  DDME2-Y & 2.12(2.62) & 11.73(13.59) & 1.54(1.50) & (1.23) & 13.20(12.99) \\
  NL3$\omega \rho$ & 2.77(3.28) & 12.99(15.88) & 1.14(0.83) & (1.47) & 13.74(13.58) \\
  NL3$\omega \rho$-Y & 2.35(2.90) & 12.58(14.77) & 1.31(1.21) & (1.35) & 13.74(13.56) \\
  SQM2 & 2.54(2.96) & 14.68(15.77) & 1.00(1.79) & (1.59) & 13.96(13.96) \\
  POL2 & 1.72(2.07) & 11.79(12.15) & 1.80(2.52) & (1.02) & 15.18(13.50) \\
 \noalign{\smallskip}
 \hline
\end{tabular}
\end{table*}


\section{Model setups}\label{sec:modsetup}

In the following, we focus on magnetised, stable models of static NSs with maximum magnetic fields  $B_\mathrm{max}\lesssim 10^{17}$G.
This value is much lower than the critical field strength, on the order of $\sim 10^{19}$G, set by the energy associated with the characteristic NS density \citep{Lattimer_Prakash_2007}. We use the following definition of the quadrupolar deformation:
\begin{equation}\label{eq:deform}
    e=\frac{I_{zz}-I_{xx}}{I_{zz}} ,\;
\end{equation}
where $I_{zz}$ and $I_{xx}$ are Newtonian moments of inertia. In STTs, these are computed in the E-frame, and they contain both the physical and the scalar field energy density (for details, see App.~C of \citetalias{soldateschi_2020}). It is known that in Newtonian theory \citep{wentzel_1960,ostriker_1969}, in GR \citep{frieben_equilibrium_2012,bucciantini_role_2015,pili_general_2017}, and in STTs \citepalias{soldateschi_2021}, in the limit of magnetic fields with maximum strength $B_\mathrm{max}\lesssim 10^{17}$G, the quadrupolar deformation $e$ can be approximated by the perturbative formulas:
\begin{equation}\label{eq:distcoeff}
    |e| = c_\mathrm{B} B^2_\mathrm{max} + \mathcal{O}\left( B^4_\mathrm{max} \right) ,\; |e| = c_\mathrm{H} \frac{\mathcal{H}}{W} + \mathcal{O}\left( \frac{\mathcal{H}^2}{W^2} \right) ,\;
\end{equation}
where $c_\mathrm{B}$ and $c_\mathrm{H}$ are called the `distortion coefficients', $B_\mathrm{max}$ is normalised to $10^{18}$G and the magnetic energy $\mathcal{H}$ and the gravitational binding energy $W$ are computed according to Eq.~5 in \citetalias{soldateschi_2021} and Eq.~C.11 in \citetalias{soldateschi_2020} respectively. In our STT case, the distortion coefficients are functions of the NS mass $M_\mathrm{k}$ {(see \citealt{gourgoulhon_3+1_2012} and App.~C of \citetalias{soldateschi_2020} for the definition of the Komar mass in GR and STTs, respectively)} and of the STT parameter, $\beta _0$, regulating the spontaneous scalarisation effect. {We note that the distortion coefficients are defined in terms of the absolute value of the quadrupolar deformation: thus, while purely poloidal and purely toroidal fields cause a quadrupolar deformation on the NS that is opposite in sign, the distortion coefficients are always positive-definite. As a result, the distortion coefficients of a NS endowed with a mixed field are expected to be always smaller than those of a NS endowed with a `pure' configuration.}

Following \citetalias{soldateschi_2020} and \citetalias{soldateschi_2021}, we used the well-tested \texttt{XNS} code \citep{bucciantini_general_2011,pili_axisymmetric_2014} to numerically solve the system of equations for static, axisymmetric, equilibrium models of magnetised NSs in the XCFC approximation \citep{cordero-carrion_improved_2009}. We used a 2D grid, using spherical coordinates: the radial coordinate extends over the range of $r\in [0,100]$ in dimensionless units, corresponding to a maximum range of $\sim 150$km; the angular coordinate extends over $\theta \in [0,\pi]$. The grid has 900 points in the $r$-direction, the first 600 of which are equally spaced over the range of $r\in [0,10]$ in dimensionless units, while the remaining 300 points are logarithmically spaced, meaning that $\Delta r_i/\Delta r_{i-1}=\mathrm{const}$. The angular grid is composed of 100 equally spaced points.

In computing the STT configurations, we adopted the following approach: we solved all the metric and scalar field equations in the E-frame, while all the MHD equations were computed in the J-frame, converting quantities from one frame to the other, when needed, through the conversion factors detailed in \citetalias{soldateschi_2020} - Sect.~2.2.

As we discussed previously, if  modifications to the internal NS structure are needed, it is worth also investigating the STT with $|\beta _0| \gtrsim 4.3$, as representative of the effects of  screened scalar fields. We chose $\alpha _0 = -2\times 10^{-4}$ and $\beta _0 \in \{ -6,-5.75,-5.5,-5 \}$. Such low values are chosen to both highlight the effects of scalarisation and to show its effects at the edge of the permitted parameter space, keeping in mind that our results hold also for scalar fields with a mass such that their screening radius is larger than the NS radius.

Our models are computed for purely poloidal and purely toroidal magnetic field configurations, the details of which are explained in Sect.~\ref{sec:grstt}. The choice of these simple, `pure' magnetic configurations, while simplifying the computations, also provides us with the results for extremal magnetic configurations: a poloidal magnetic field affects the NS deformation in an opposite way with respect to a toroidal field; as such, we expect configurations with mixed fields to show a deformation which is{, in absolute value, smaller} than those obtained with purely poloidal and toroidal configurations.

To compute the results shown in this paper, we first used the full \texttt{XNS} code to compute roughly $65000$ numerical models with stronger magnetic fields and then we interpolated the results according to the approximations in Eq.~\ref{eq:distcoeff}. The large number of models used allows us to limit the errors introduced in the interpolation process. We studied only those configurations belonging to the stable branch of the mass-density diagram, that is, with a central density, $\rho _\mathrm{c}$, lower than that of the maximum mass model. Many results shown in Sect.~\ref{sec:results} and, in particular, the quasi-universal relations, were obtained through a `principal component analysis' (PCA). Briefly, PCA is a dimensionality reduction technique used to find correlations among a given set of data. In particular, the PCA algorithm computes the $D$ `principal components' of some input $D$-dimensional data, namely, the hyperplanes that best fit the input data and are orthogonal to each other. While the first principal component is the single hyperplane that maximises the variance of data projected onto it, the last ($D$-th) principal component is the hyperplane around which data is spread out the least. For this reason, the quasi-universal relations that we show in the following are precisely the equations of the last PCA component, that is, the hyperplane that best fits the input data. In providing the input data to the PCA algorithm, we selected only those configurations with a mass, $M_\mathrm{k}>1$M$_\odot$, both in GR and in STT, in order to analyse only models with a realistic mass value. Moreover, in the STT case, we selected scalar charges of $Q_\mathrm{s}>0.4$M$_\odot$; this is done because, as we go on to show, in STTs, we are mostly interested in deviation from GR. In this respect models with small scalar charges, especially in the small region where spontaneous scalarisation abruptly develops, tend to be less accurate. Let us remark, however, that in terms of the total quadrupolar deformation (excluding such low values of the scalar charge from the PCA) does not change our results in an appreciable way: when the value of the scalar charge approaches zero, the deviation from GR becomes increasingly negligible, and the GR quasi-universal relations hold. In practice, large relative errors translate into small absolute ones.
\begin{figure*}[!ht]
  \centering
  \subfloat[][]{\includegraphics[width=.49\textwidth]{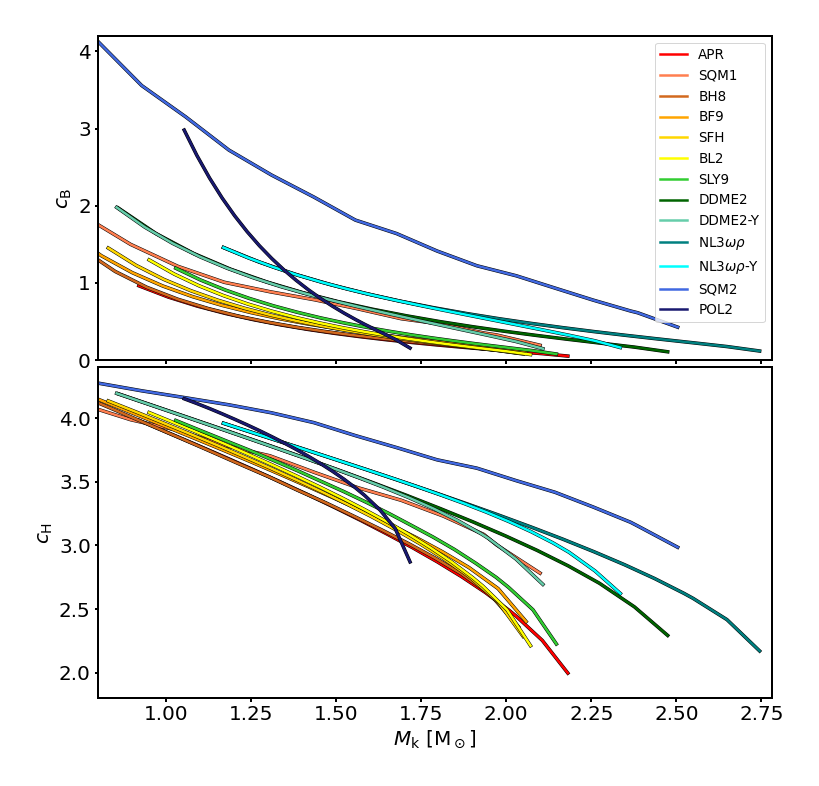}}\quad
  \subfloat[][]{\includegraphics[width=.49\textwidth]{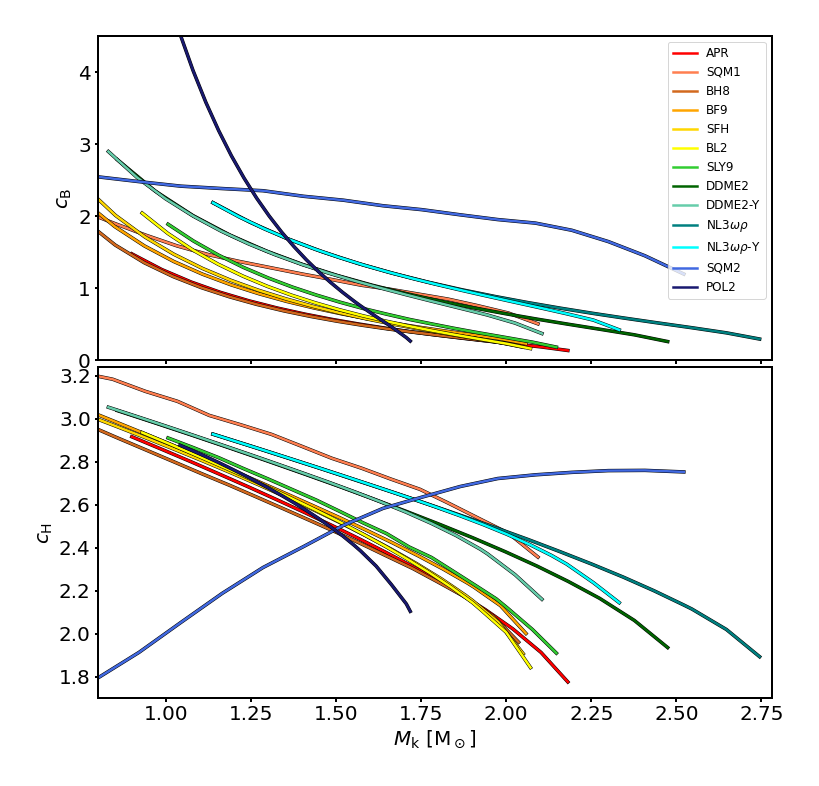}}\\
  \caption{Distortion coefficients $c_\mathrm{B}$ and $c_\mathrm{H}$, calculated according to Eq.~\ref{eq:distcoeff}, as functions of the Komar mass $M_\mathrm{k}$ for purely poloidal [(a) panel] and purely toroidal [(b) panel] magnetic fields, in GR. The EoS are colour-coded, and ordered in the legend, according to the compactness $C=M_\mathrm{k}/R_\mathrm{c}$ calculated at $M_\mathrm{k}=1.4$M$_\odot$ in GR: red for the highest compactness and blue for the lowest compactness.}
  \label{fig:cbh_mk_gr}
\end{figure*}
\section{Results}\label{sec:results}
In this section, we first describe how different EoS affect the magnetic structure of our NS models. Then we detail the results obtained on the distortion coefficients using all EoS described in Sect.~\ref{sec:eos} and show the quasi-universal relations we found, considering only viable EoS that describe standard NSs, that is, excluding SQM1, SQM2, and POL2. Afterwards, we comment on how these results apply to polytropes. We perform this analysis first for GR and then for STTs. Finally, we comment on how our results apply to models of strange quark stars.

Our poloidal models are characterised by a magnetic field whose magnitude is always maximum at the star centre, vanishing in an equatorial current ring located typically at $\sim 60-70 \%$ of the NS radius, followed by smaller secondary maximum just underneath the surface. At the surface, the maximum is always reached at the pole and typically is $\sim 20-25\%$ of the central value. On the other hand, the models endowed with a purely toroidal magnetic field possess a simpler magnetic structure: the magnitude of the magnetic field is zero on the axis, reaches a maximum inside the NS, typically at $\sim 40-60\%$ of the NS radius, and then decreases to zero at the surface. The exact profile of the magnetic field components depends on the EoS. In fact, while the general behaviour of the magnetic field is in line with that just described above, the particular composition of the NS causes a shift in the location of the maxima and in the smoothness of the magnetic field profile. As described in Sect.~\ref{sec:eos}, the BH8 EoS contains a pion condensate, leading to the appearance of a jump in density. As a consequence, the magnetic field of purely toroidal models display a jump in the magnetic field magnitude at the same location where the condensate appears. On the other hand, purely poloidal models show no sudden change in the magnetic profile (the jump is present in the current distribution). In the case of models described by the SQM1 and SQM2 EoS, for purely poloidal magnetic fields, the equatorial ring where the magnetic field vanishes is located at a radius, which is $\sim 10\%$ higher than for the standard EoS. In the case of purely toroidal models, this shift is such that the magnetic profile is abruptly truncated at the star surface; in some cases, the magnetic field strength monotonically rises all the way to the surface, where it reaches its maximum and then jumps to zero. As we comment in Sect.~\ref{subsec:sqm}, this behaviour leads us to consider these models as not true equilibria, since a non-vanishing Lorentz force at the NS surface remains unbalanced. As for the hyperonic EoS, the appearance of hyperons causes an increase in the maximum magnitude of the magnetic field with respect to models of the same central density computed with the corresponding non-hyperonic EoS. Finally, we note that although the profiles of the density and of the magnetic field are affected by the EoS (condensates, appearance of new particles, etc..) as we just described, all the integrated, global quantities, such as the mass, radius, magnetic energy, scalar charges, or quadrupolar deformation of the NS, show no sign of discontinuities or jumps.
\subsection{General relativity}
\begin{figure}[!ht]
  \centering
  \includegraphics[width=\columnwidth]{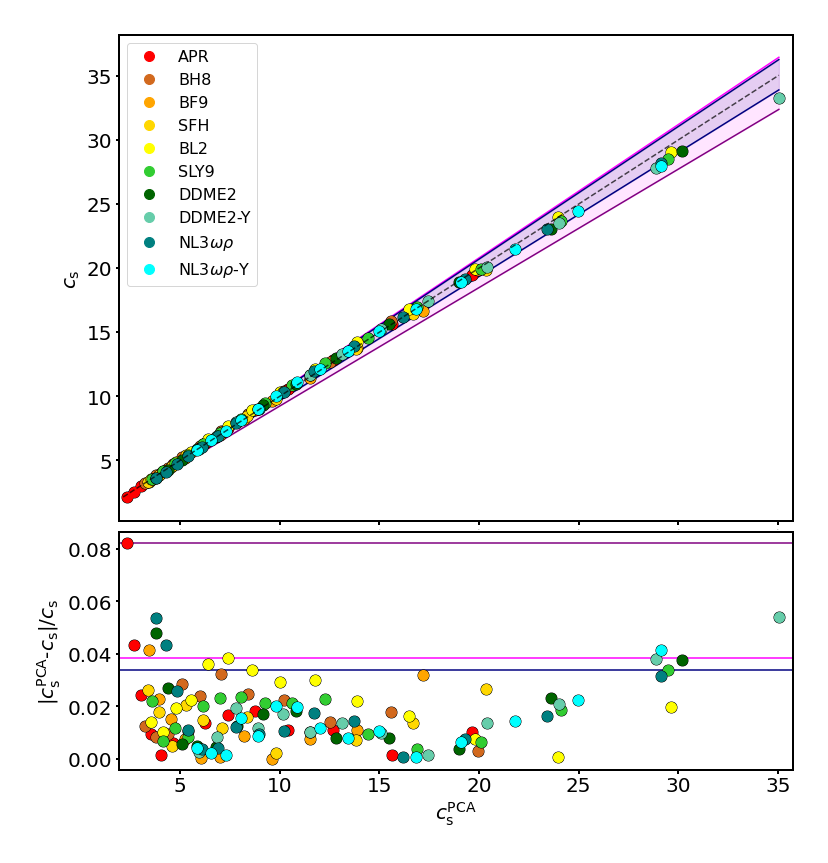}
  \caption{Distortion coefficient $c_\mathrm{s}$, calculated according to Eq.~\ref{eq:distcoeff_cs} in GR, versus its approximation $c^\mathrm{PCA}_\mathrm{s}$ calculated with the quasi-universal relation in Eq.~\ref{eq:cs_pca_gr} (top plot). The corresponding relative deviation from the PCA is given in the bottom plot. The dashed line is $c_\mathrm{B,H} = c^\mathrm{PCA}_\mathrm{B,H}$. The magenta shaded area comprises all data points and the purple and magenta lines represent the upper and lower bounds of Eq.~\ref{eq:cs_pca_gr}. The dark blue lines bounding the shaded blue area mark the 90th percentile error region. The EoS are colour-coded, and ordered in the legend, according to the compactness $C=M_\mathrm{k}/R_\mathrm{c}$ calculated at $M_\mathrm{k}=1.4$M$_\odot$ in GR: red for the highest compactness and blue for the lowest compactness.}
  \label{fig:pca_cs_gr}
\end{figure}

The distortion coefficients, $c_\mathrm{B}$ and $c_\mathrm{H}$, for our NS models, calculated in GR, are shown in Fig.~\ref{fig:cbh_mk_gr} as functions of the NS Komar mass, $M_\mathrm{k}$\footnote{The distortion coefficients, $c_\mathrm{B}$ and $c_\mathrm{H}$, are shown separately for each EoS described in Sect.~\ref{sec:eos}, both in GR and in STTs, in the supplementary materials: \url{https://doi.org/10.5281/zenodo.5336222}.}. Panel (a) refers to models endowed with a purely poloidal magnetic field, while panel (b) referes to purely toroidal magnetic fields. The EoS are colour-coded according to the compactness $C=M_\mathrm{k}/R_\mathrm{c}$, calculated at $M_\mathrm{k}=1.4$M$_\odot$ in GR: red for the highest compactness and blue for the lowest compactness. We can see that sequences for both coefficients are only roughly ordered according to the compactness of the EoS: models with the same mass have lower distortion coefficients for more compact EoS only on average, the main exceptions to this rule being the SQM1, SQM2 and POL2 EoS. This is especially true for $c_\mathrm{H}$ in the toroidal case, where the POL2 EoS (the least compact one) reaches a lower distortion coefficient than the APR EoS  (the most compact one) close to its maximum-mass model and the SQM2 EoS displays a completely different behaviour. We note that while it is to be expected that more compact EoS have a lower deformation, the particular definition of compactness we use (calculated for 1.4M$_\odot$ models) clearly impacts the ordering of the EoS, as can be seen from the mass-radius relations in Fig.~\ref{fig:mrrel}. If we exclude the SQM1, SQM2, and POL2 EoS, we see that all sequences are more closely packed: the relative difference, in the poloidal (toroidal) cases, $2(c^+_\mathrm{B}-c^-_\mathrm{B})/(c^+_\mathrm{B}+c^-_\mathrm{B})$ between the uppermost sequence (+ superscript, for the NL3$\omega \rho$ EoS) and the lowermost sequence (- superscript, for the APR EoS) in the poloidal (toroidal) case is $\sim 1.26$($\sim 1.14$) at $M_\mathrm{k}\sim 2.0$M$_\odot$ and $\sim 0.83$($\sim 0.79$) at $M_\mathrm{k}\sim 1.2$M$_\odot$, indicating that above 1.7M$_\odot$, various EoS can differ by order unity. For $c_\mathrm{H}$, in the poloidal (toroidal) case, the difference $2(c^+_\mathrm{H}-c^-_\mathrm{H})/(c^+_\mathrm{H}+c^-_\mathrm{H})$ ranges from $\sim 0.25$ ($\sim 0.19$) at $M_\mathrm{k}\sim 2.0$M$_\odot$, to $\sim 0.07$ ($\sim 0.07$) at $M_\mathrm{k}\sim 1.2$M$_\odot$ . These numbers show that while $c_\mathrm{B}$ vastly differs between the sequences for the APR and the NL3$\omega \rho$ EoS, the value of $c_\mathrm{H}$ remains almost constant across the whole mass range, changing at most by a factor of $\sim 1.5$.
Since the SQM1, SQM2, and POL2 EoS display such a different behaviour, for the moment, we focus only on those EoS describing standard NS (thus excluding SQM1 and SQM2) allowed by observations (thus excluding POL2), which we refer to as `standard EoS'.
\\\\
\begin{figure*}[!htp]
  \centering
  \subfloat[][]{\includegraphics[width=.49\textwidth]{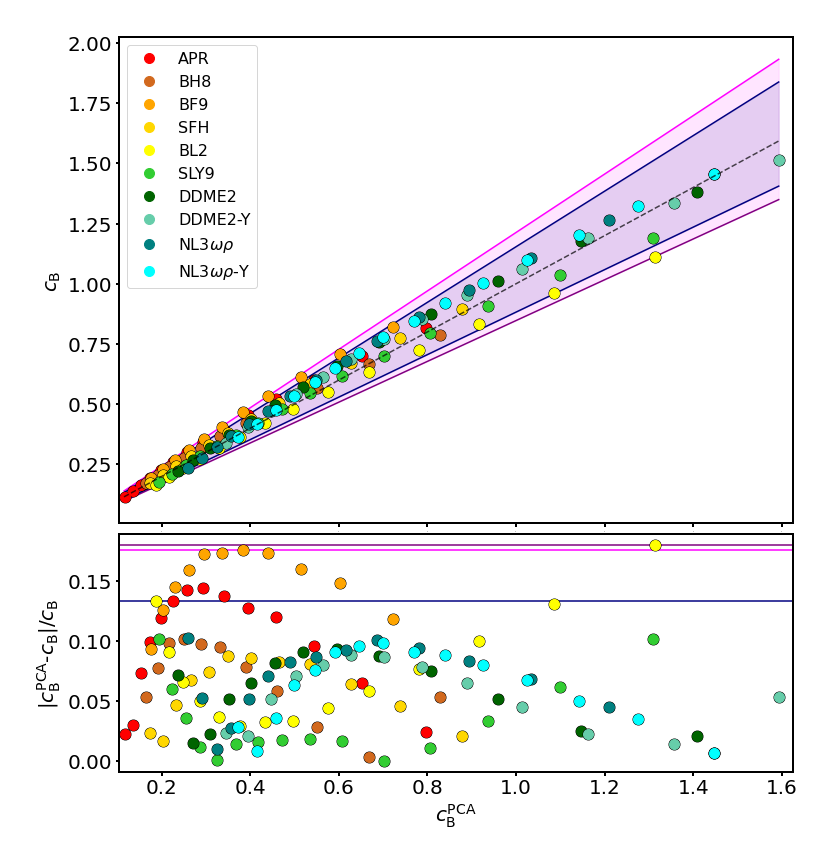}}\quad
  \subfloat[][]{\includegraphics[width=.49\textwidth]{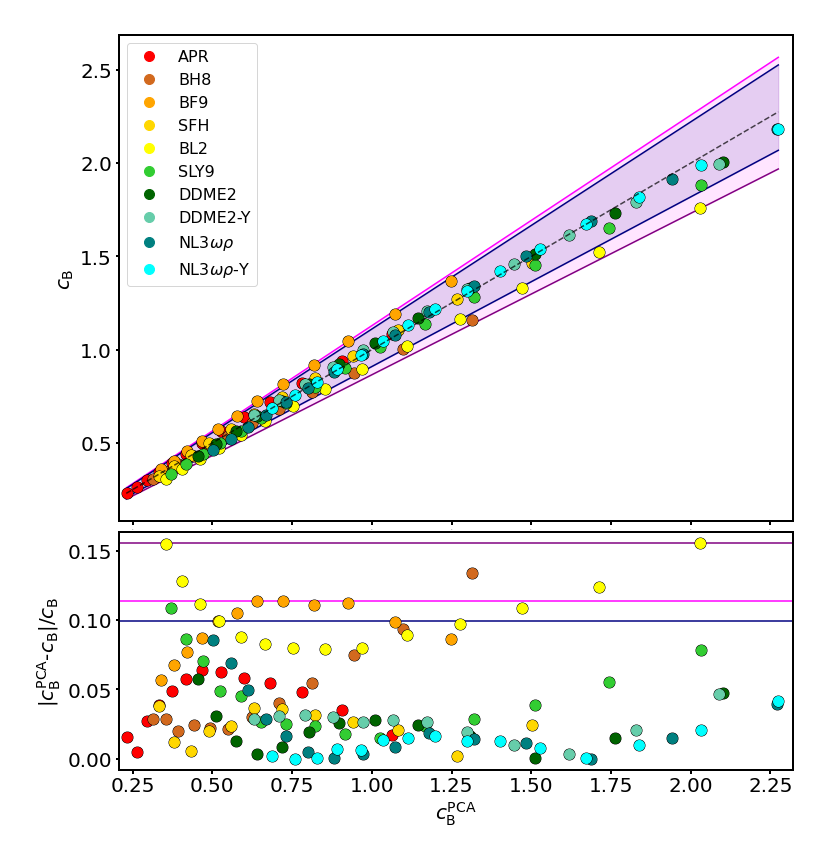}}\\
  \subfloat[][]{\includegraphics[width=.49\textwidth]{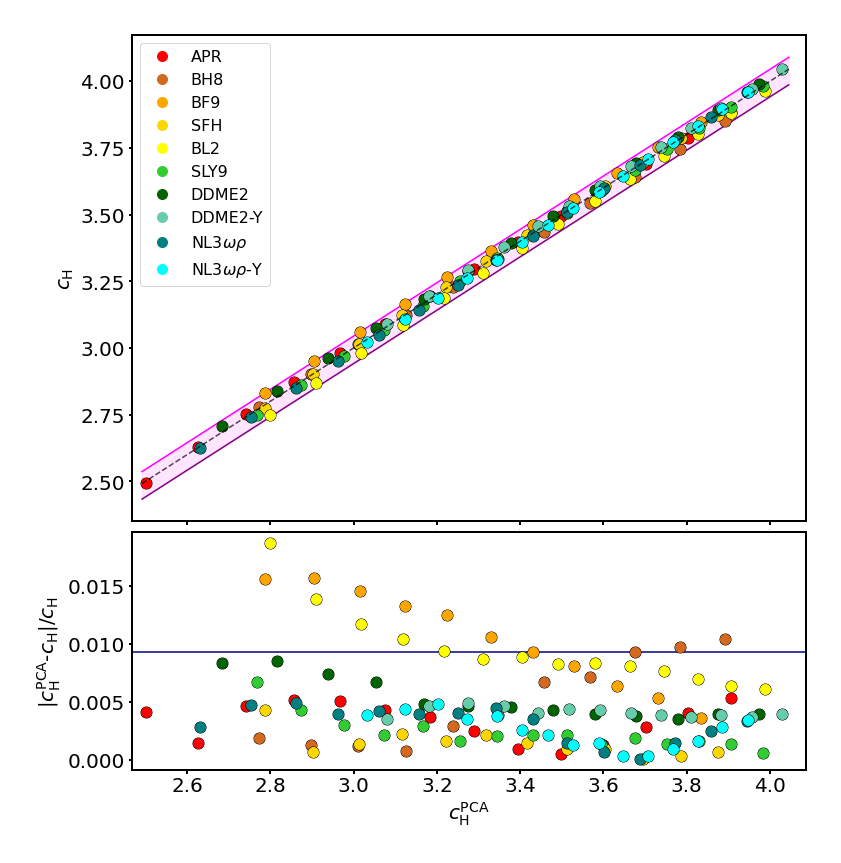}}\quad
  \subfloat[][]{\includegraphics[width=.49\textwidth]{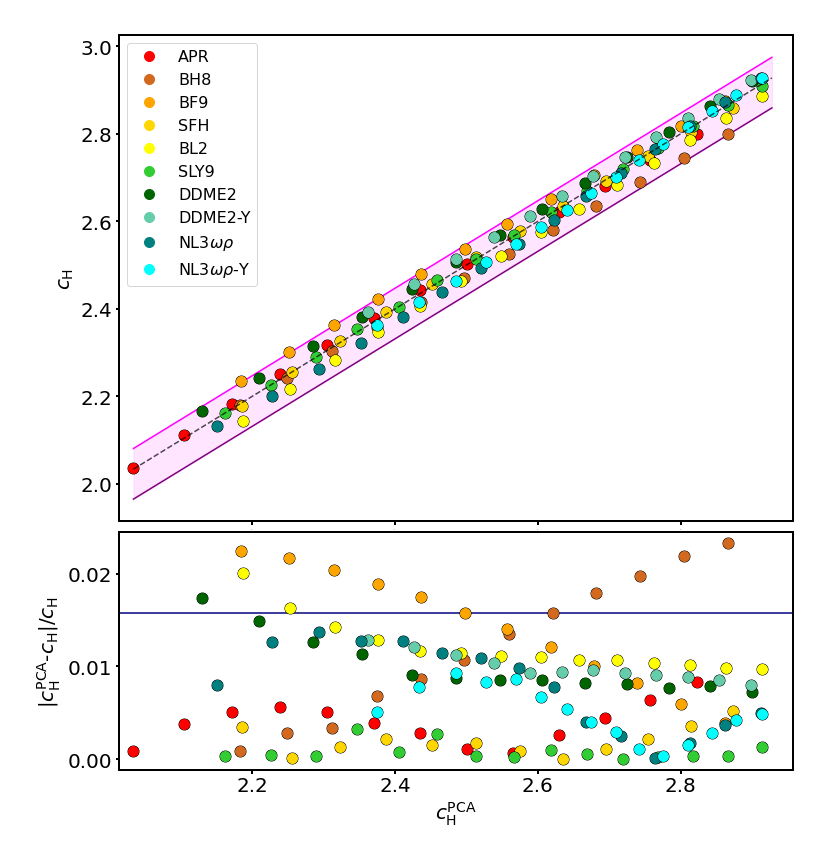}}
  \caption{Distortion coefficients $c_\mathrm{B}$ (top panels) and $c_\mathrm{H}$ (bottom panels), calculated according to Eq.~\ref{eq:distcoeff} in GR, versus their approximations $c^\mathrm{PCA}_\mathrm{B}$ and $c^\mathrm{PCA}_\mathrm{H}$, calculated with the quasi-universal relations in Eqs.~\ref{eq:cb_pca_gr}-\ref{eq:ch_pca_gr} (top plot in each panel). The corresponding relative deviations from the PCA are given in the bottom plot in each panel. Panels (a) and (c) refer to purely poloidal magnetic fields; panels (b) and (d) refer to purely toroidal magnetic fields. The dashed black line is $c_\mathrm{B,H} = c^\mathrm{PCA}_\mathrm{B,H}$. The magenta shaded area comprises all data points and the purple and magenta lines represent the upper and lower bounds of Eqs.~\ref{eq:cb_pca_gr}-\ref{eq:ch_pca_gr}. The dark blue lines bounding the shaded blue area mark the 90th percentile error region. The EoS are colour-coded, and ordered in the legend, according to the compactness $C=M_\mathrm{k}/R_\mathrm{c}$ calculated at $M_\mathrm{k}=1.4$M$_\odot$ in GR: red for the highest compactness and blue for the lowest compactness.}
  \label{fig:pca_cbh_gr}
\end{figure*}
Given the similarity between the distortion coefficients for all standard EoS when plotted against the Komar mass, it is reasonable to wonder whether adding the dependence on other variables could further reduce the spread. We thus chose to consider the dependence also on another potentially observable quantity, namely the circumferential radius $R_\mathrm{c}$, and adopted a PCA algorithm to find the best-fit relation between $c_\mathrm{B,H}$, $M_\mathrm{k}$, and $R_\mathrm{c}$. We found that these formulas approximate $c_\mathrm{B,H}$ to a satisfying level of accuracy for all standard EoS:
\begin{equation}\label{eq:cb_pca_gr}
        c^\mathrm{PCA}_\mathrm{B} =
            \begin{cases}
                0.13^{+0.03}_{-0.02} R_{10}^{5.45} M_{1.6}^{-2.41} \;{\rm for\; poloidal,}\\\\
                0.25^{+0.03}_{-0.03} R_{10}^{5.03} M_{1.6}^{-2.07} \;{\rm for\; toroidal,}
            \end{cases}
\end{equation}
\begin{align}\label{eq:ch_pca_gr}
        c^\mathrm{PCA}_\mathrm{H} =
            \begin{cases}
                5.77^{+0.04}_{-0.06} - 0.77 R_{10} - 4.14 M_{1.6} - 0.27 M_{1.6}^2 +\\ \;\;+ 0.07 R_{10}^2 + 2.28 M_{1.6} R_{10} \;{\rm for\; poloidal,}\\\\
                7.02^{+0.05}_{-0.07} - 5.22 R_{10} - 2.76 M_{1.6} - 0.12 M_{1.6}^2 +\\ \;\;+ 1.92 R_{10}^2 + 1.51 M_{1.6} R_{10} \;{\rm for\; toroidal,}
            \end{cases}
\end{align}
where $R_{10} = R_\mathrm{c}/10\mathrm{km}$ and $M_{1.6} = M_\mathrm{k}/1.6$M$_\odot$. We refer to these formulas, as well as the analogue ones described in the following, as `quasi-universal relations', since they hold for all standard EoS we considered. The distortion coefficients computed using formulas from Eqs.~\ref{eq:cb_pca_gr}-\ref{eq:ch_pca_gr} are plotted against their `real' value, computed with formulas Eq.~\ref{eq:distcoeff}, as shown in in Fig.~\ref{fig:pca_cbh_gr} (top plot in each panel) for purely poloidal (left panels) and purely toroidal (right panels) magnetic field configurations. In each panel, the bottom plot displays the relative error $|c_\mathrm{B,H}^\mathrm{PCA}-c_\mathrm{B,H}|/c_\mathrm{B,H}$ committed when using the quasi-universal relations to approximate the distortion coefficients. The dashed black line is a reference $c_\mathrm{B,H} = c^\mathrm{PCA}_\mathrm{B,H}$ bisecting line, which stands for a perfect approximation. The superscripts and subscripts in the first coefficient of Eqs.~\ref{eq:cb_pca_gr}-\ref{eq:ch_pca_gr} are the values that define the purple and magenta lines bounding the magenta shaded area in Fig.~\ref{fig:pca_cbh_gr}, top plots in each panel. The dark blue lines, which bound the shaded blue area in the plots of $c_\mathrm{B}$, mark the 90th percentile of the relative errors (the bounds containing 90\% of the results). The corresponding values of these errors are showed with lines of the same colour in the bottom plots of each panel for $c_\mathrm{B}$. In the case of $c_\mathrm{H}$, given the form of the quasi-universal relation Eq.~\ref{eq:ch_pca_gr}, the bounds in the top plots do not correspond to a unique constant value of the relative error, thus we omit them in the bottom plots; likewise, we omit the 90th percentile line in the top plots. We note that the magenta regions contain all the points, and, in this sense, the bounds in Eqs.~\ref{eq:cb_pca_gr}-\ref{eq:ch_pca_gr} represent the maximum spread of the results; however, in general, typical deviations with respect to the PCA approximations are about 3/4 to 1/2 of those values. We see that the quasi-universal relations for $c_\mathrm{B}$, in the poloidal case [(a) panel], hold with a maximum relative error of $\sim 17\%$, and the 90th percentile stands at $\sim 14\%$; in the toroidal case [(b) panel], they are $\sim 16\%$ and $\sim 10\%,$ respectively. The approximation for $c_\mathrm{H}$ is much more accurate, with a maximum relative error of $\sim 2\%$ in both magnetic configurations, and mostly under $\sim 1\%$ in the poloidal case [(c) panel] and under $\sim 1.5\%$ in the toroidal case [(d) panel]. We see no dependence of the deviation from the PCA results on the compactness of the EoS. We note that the coefficients for $c^\mathrm{PCA}_\mathrm{H}$ in Eq.~\ref{eq:ch_pca_gr} in the toroidal case can be used also in the poloidal case, but in this case, the 90th percentile relative error increases to $\sim 27\%$. However, similarly to what we found in \citetalias{soldateschi_2021} for $c_\mathrm{H}$, performing $c^\mathrm{PCA}_\mathrm{H} \rightarrow 5/3c^\mathrm{PCA}_\mathrm{H}-0.9$ allows one to use the toroidal coefficients in the poloidal case with a $\sim 2\%$ error.
\\\\
The distortion coefficients as defined in Eq.~\ref{eq:distcoeff} contain quantities that are not directly accessible by observations, since they require knowledge of the details of the internal structure and magnetic field geometry of NS. In this respect,  the quasi-universal relations Eqs.~\ref{eq:cb_pca_gr}-\ref{eq:ch_pca_gr} may be used to get  information on the internal structure of the magnetic field, as we discuss in Sect.~\ref{sec:constr}. However, from an observational prospective,  it is useful to introduce another distortion coefficient, defined using a quantity that may be observed:
\begin{equation}\label{eq:distcoeff_cs}
    |e| = c_\mathrm{s} B^2_\mathrm{s} + \mathcal{O}\left( B^4_\mathrm{s} \right) \; ,
\end{equation}
where $B_\mathrm{s}$ is the magnetic field calculated at the pole of the NS, at the surface, normalised to $10^{18}$G. Obviously this coefficient is defined only for configurations endowed with a poloidal magnetic field (the toroidal one being hidden under the surface). As it was done for $c_\mathrm{B}$ and $c_\mathrm{H}$, we performed a PCA and found the corresponding quasi-universal relation:
\begin{equation}\label{eq:cs_pca_gr}
    \begin{split}
        c^\mathrm{PCA}_\mathrm{s} &= 2.97^{+0.12}_{-0.23} R_{10}^{4.61} M_{1.6}^{-2.80} \; .
    \end{split}
\end{equation}
The distortion coefficient $c_\mathrm{s}$ computed using formula Eq.~\ref{eq:cs_pca_gr} is plotted against its `real' value, computed with formula Eq.~\ref{eq:distcoeff_cs}, in Fig.~\ref{fig:pca_cs_gr} (top plot), along with the relative error of the approximation (bottom panel). We note that the values of $c_\mathrm{s}$ are roughly one order of magnitude higher than those of $c_\mathrm{B}$ because the magnetic field at the surface is lower than the internal one, while the normalisation used is the same for both coefficients. We see that the approximation $c^\mathrm{PCA}_\mathrm{s}$ holds to a satisfying accuracy, with a maximum relative error of $~8\%$, but mostly concentrated under $4\%$.
\\\\
As can be seen from Fig.~\ref{fig:cbh_mk_gr}, applying the PCA derived from standard EoS, Eqs.~\ref{eq:cb_pca_gr}-\ref{eq:ch_pca_gr}-\ref{eq:cs_pca_gr}, to the model computed with the POL2 EoS leads to a large errors: the PCA approximation is larger by a factor of $\sim 1.9$ for $c_\mathrm{B}$ both in the poloidal (at 1.05M$_\odot$) and toroidal (at 1.55M$_\odot$) case, and a factor of $\sim 1.8$ larger at 1.4M$_\odot$; instead, the relative deviation reaches $\sim 7\%$ at 1.56M$_\odot$ $(\sim 20\%$ at 1.04M$_\odot$ $)$ for $c_\mathrm{H}$ in the poloidal (toroidal) case, while it is $\sim 6\% (\sim 17\%)$ for 1.4M$_\odot$ models in the poloidal (toroidal) case; similarly, the maximum error for $c_\mathrm{s}$ is $\sim 20\%$ at 1.05M$_\odot$, while it is $\sim 2\%$ at 1.4M$_\odot$.

\subsection{Scalar-tensor theories}
\begin{figure*}[!htp]
  \centering
  \subfloat[][]{\includegraphics[width=.49\textwidth]{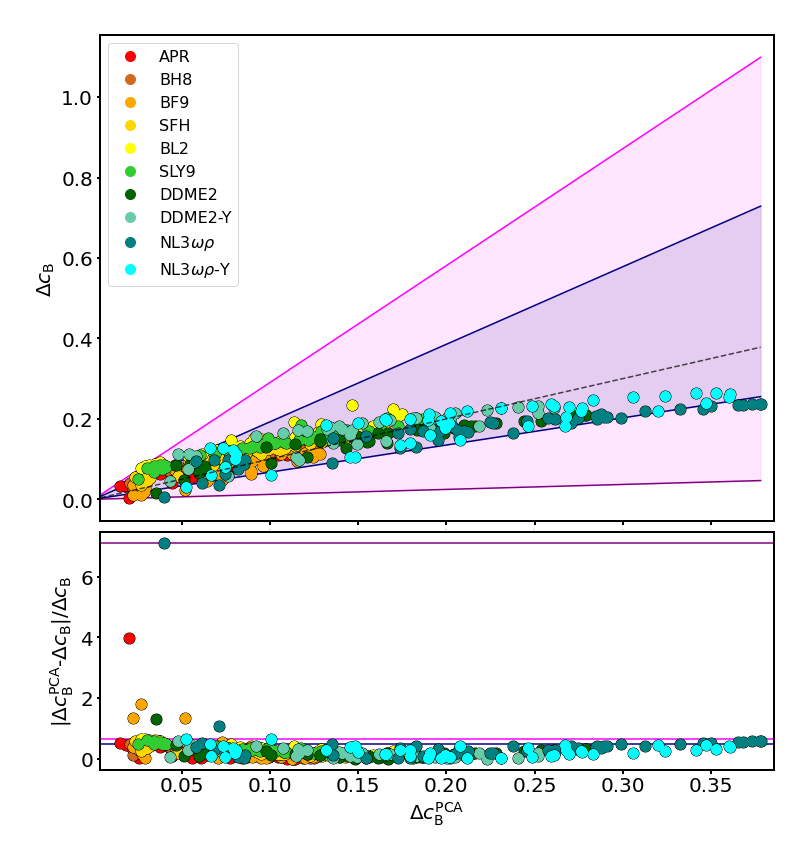}}\quad
  \subfloat[][]{\includegraphics[width=.49\textwidth]{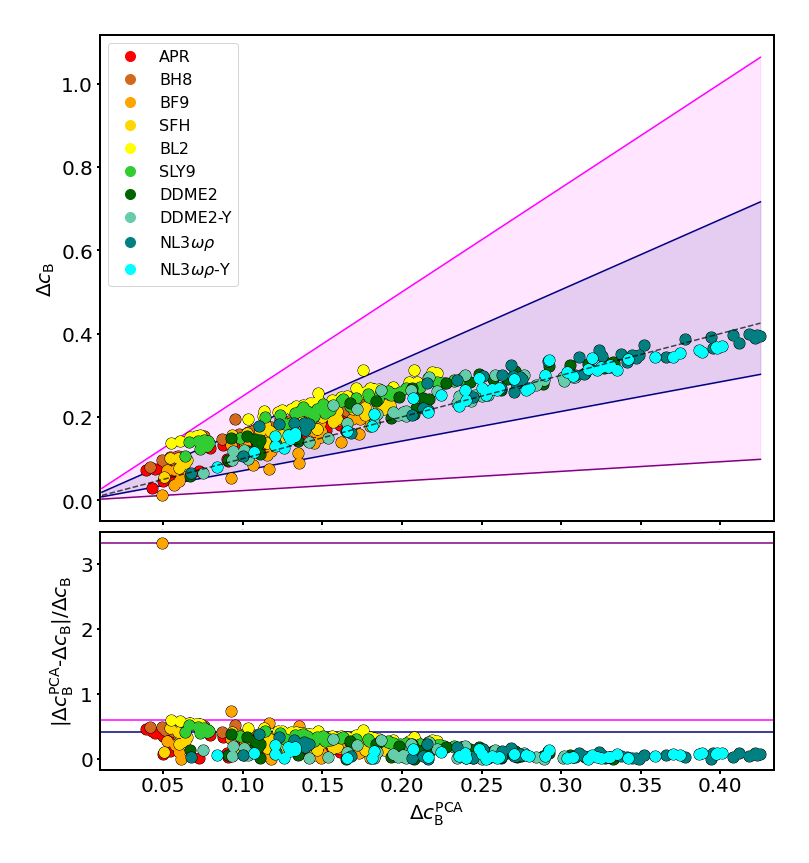}}\\
  \subfloat[][]{\includegraphics[width=.49\textwidth]{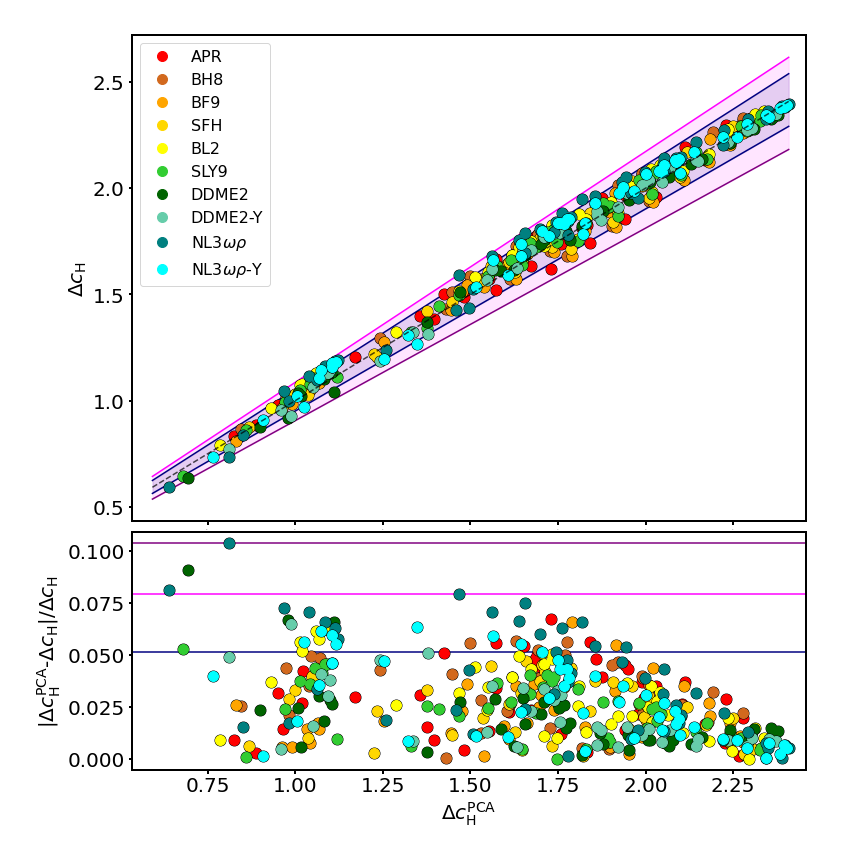}}\quad
  \subfloat[][]{\includegraphics[width=.49\textwidth]{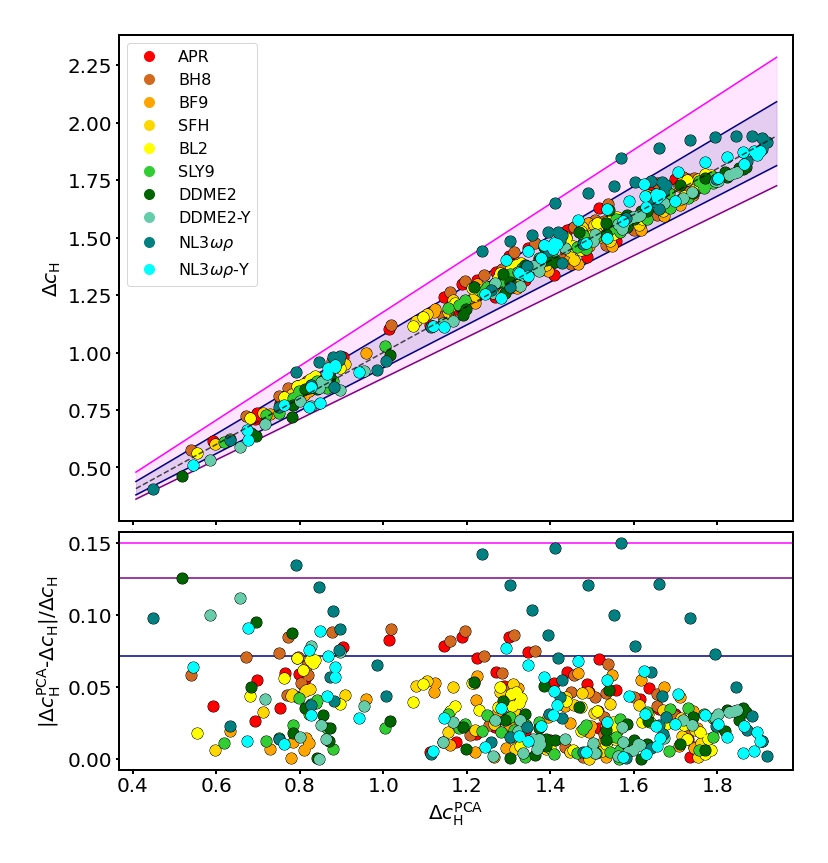}}
  \caption{Differences $\Delta c_\mathrm{B}$ (top panels) and $\Delta c_\mathrm{H}$ (bottom panels) between the distortion coefficients $c_\mathrm{B}$ and $c_\mathrm{H}$, calculated according to Eq.~\ref{eq:distcoeff} for STTs, with $\beta _0~\in~\{ -6,-5.75,-5.5,-5 \}$, and the GR quasi-universal relations in Eq.~\ref{eq:cb_pca_gr}-\ref{eq:ch_pca_gr}, respectively. These are plotted versus  $c^\mathrm{PCA}_\mathrm{B}$ and $c^\mathrm{PCA}_\mathrm{H}$and  calculated as in Eqs.~\ref{eq:cb_pca_stt}-\ref{eq:ch_pca_stt} (top plot in each panel). The corresponding relative deviations from the PCA are given in the bottom plot in each panel. Panels (a) and (c) refer to purely poloidal magnetic fields; panels (b) and (d) refer to purely toroidal magnetic fields. The dashed line is $\Delta c_\mathrm{B,H} = \Delta c^\mathrm{PCA}_\mathrm{B,H}$. The magenta shaded area comprises all data points and the purple and magenta lines represent the upper and lower bounds of Eqs.~\ref{eq:cb_pca_stt}-\ref{eq:ch_pca_stt}; the dark blue lines bounding the shaded blue area mark the 90th percentile error region. The EoS are colour-coded, and ordered in the legend, according to the compactness $C=M_\mathrm{k}/R_\mathrm{c}$  at $M_\mathrm{k}=1.4$M$_\odot$ in GR: red for the highest compactness and blue for the lowest.}
  \label{fig:pca_cbh_stt}
\end{figure*}

\begin{figure}[!htp]
  \centering
  \includegraphics[width=\columnwidth]{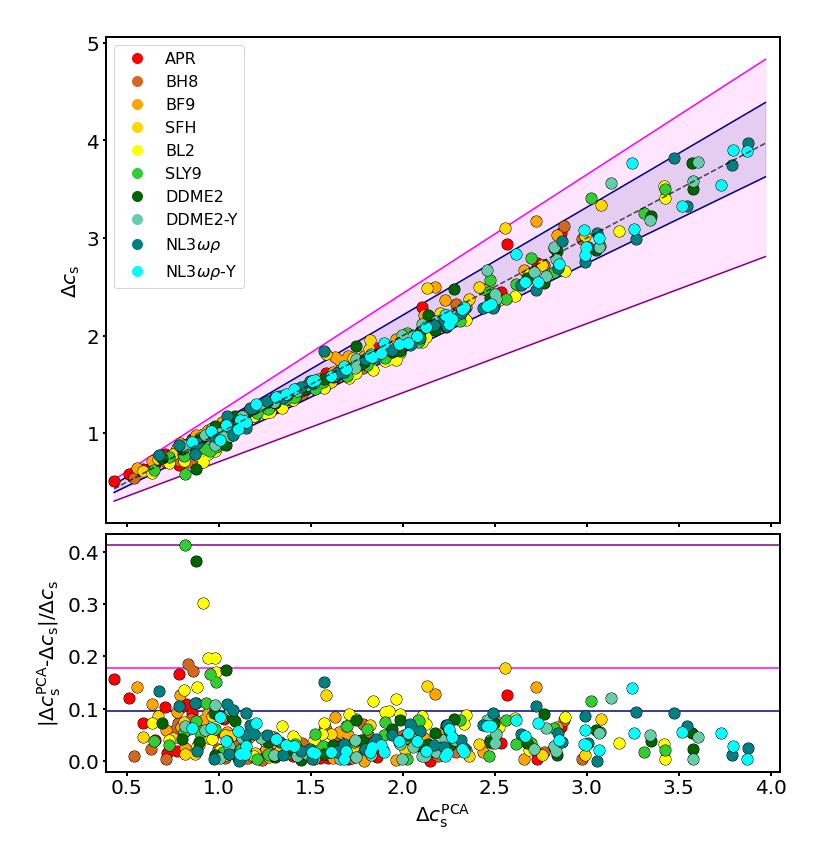}
  \caption{Difference, $\Delta c_\mathrm{s}$, between the distortion coefficient $c_\mathrm{s}$, calculated according to Eq.~\ref{eq:distcoeff_cs} for STTs with $\beta _0~\in~\{ -6,-5.75,-5.5,-5 \}$, and the GR quasi-universal relation in Eq.~\ref{eq:cs_pca_gr}. This is plotted versus its approximation $c^\mathrm{PCA}_\mathrm{s}$, calculated with the quasi-universal relation in Eqs.~\ref{eq:cs_pca_stt} (top plot). The corresponding relative deviation from the PCA is given in the bottom plot. The dashed line is $\Delta c_\mathrm{s} = \Delta c^\mathrm{PCA}_\mathrm{s}$. The magenta shaded area comprises all data points and the purple and magenta lines represent the upper and lower bounds of Eq.~\ref{eq:cs_pca_stt}. The dark blue lines bounding the shaded blue area mark the 90th percentile error region. The EoS are colour-coded, and ordered in the legend, according to the compactness $C=M_\mathrm{k}/R_\mathrm{c}$ calculated at $M_\mathrm{k}=1.4$M$_\odot$ in GR: red for the highest compactness and blue for the lowest compactness.}
  \label{fig:pca_cs_stt}
\end{figure}

In the case of STTs, we find quasi-universal relations for $\Delta c_\mathrm{B} = |c_\mathrm{B} - c^\mathrm{GR}_\mathrm{B}|$, $\Delta c_\mathrm{H} = |c_\mathrm{H} - c^\mathrm{GR}_\mathrm{H}|$ and $\Delta c_\mathrm{s} = |c_\mathrm{s} - c^\mathrm{GR}_\mathrm{s}|$, where $c^\mathrm{GR}_\mathrm{B}, c^\mathrm{GR}_\mathrm{H}$ and $c^\mathrm{GR}_\mathrm{s}$ are the relations found in the GR case: Eqs.~\ref{eq:cb_pca_gr}-\ref{eq:ch_pca_gr}-\ref{eq:cs_pca_gr}, respectively. We chose this approach instead of approximating the bare distortion coefficients for two reasons: on the one hand, since we already found satisfying approximations in GR, it makes sense to focus only on the difference given by scalarisation; on the other hand, this allows us to exclude the few models with a low scalar charge, which are located in the mass-radius diagram in Fig.~\ref{fig:mrrel} (bottom panel), close to the sharp onset of scalarisation, and which may not be as accurately computed as the rest of the sequence. In computing the PCA in the STT case, we considered also the dependence of the distortion coefficients on the scalar charge, $Q_\mathrm{s}$ {(see App.~C of \citetalias{soldateschi_2020} for the definition)}. We found the following quasi-universal relations:
\begin{equation}\label{eq:cb_pca_stt}
        \Delta c^\mathrm{PCA}_\mathrm{B} =
            \begin{cases}
                0.03^{+0.05}_{-0.03} R_{10}^{8.23} M_{1.6}^{-5.08} Q_{1}^{2.60} \;{\rm for\; poloidal,}\\\\
                0.06^{+0.09}_{-0.05} R_{10}^{5.96} M_{1.6}^{-3.52} Q_{1}^{1.95} \;{\rm for\; toroidal,}
            \end{cases}
\end{equation}
\begin{equation}\label{eq:ch_pca_stt}
        \Delta c^\mathrm{PCA}_\mathrm{H} =
            \begin{cases}
                1.96^{+0.17}_{-0.18} R_{10}^{0.72} M_{1.6}^{-1.96} Q_{1}^{1.54} \;{\rm for\; poloidal,}\\\\
                1.49^{+0.26}_{-0.17} R_{10}^{0.75} M_{1.6}^{-1.81} Q_{1}^{1.55} \;{\rm for\; toroidal,}
            \end{cases}
\end{equation}
\begin{equation}\label{eq:cs_pca_stt}
        \Delta c^\mathrm{PCA}_\mathrm{s} =
                0.92^{+0.20}_{-0.27} R_{10}^{4.77} M_{1.6}^{-4.50} Q_{1}^{1.71} ,
\end{equation}
where $Q_{1}$ is $Q_\mathrm{s}$ normalised to 1M$_\odot$. The quasi-universal relations Eqs.~\ref{eq:cb_pca_stt}-\ref{eq:ch_pca_stt} are plotted against the corresponding value $\Delta c_\mathrm{B}$ and $\Delta c_\mathrm{H}$, computed using formulas Eq.~\ref{eq:distcoeff}, in Fig.~\ref{fig:pca_cbh_stt} (top plot in each panel) for purely poloidal (left panels) and purely toroidal (right panels) magnetic field configurations. Instead, Eq.~\ref{eq:cs_pca_stt} is plotted against the corresponding value $\Delta c_\mathrm{s}$, computed using formula Eq.~\ref{eq:distcoeff_cs}, in Fig.~\ref{fig:pca_cs_stt} (top plot). In each figure and panel, the bottom plot displays the relative error of the quasi-universal relations. The dashed lines are a reference $\Delta c_\mathrm{B,H} = \Delta c^\mathrm{PCA}_\mathrm{B,H}$ and $\Delta c_\mathrm{s} = \Delta c^\mathrm{PCA}_\mathrm{s}$ bisecting lines, which stand for a perfect approximation.
We see from the bottom plots in each panel of Fig.s~\ref{fig:pca_cbh_stt}-\ref{fig:pca_cs_stt} that the overall relative errors are larger than in the GR case. This is to be expected for two reasons: on the one hand, we are approximating the $\Delta c_\mathrm{B,H,s}$, which, by definition, are computed on the difference with an already approximated quantity, $c^\mathrm{PCA}_\mathrm{B,H,s}$; on the other hand, especially at low values of the scalar charge (bottom-left of each top plot), the onset of scalarisation causes an abrupt change in the mass-radius relation compared to GR, inevitably decreasing the accuracy of NS models. We see that the quasi-universal relations for $\Delta c_\mathrm{B}$, both in the poloidal [(a) panel] and in the toroidal case [(b) panel], hold with a 90th percentile relative error of $\sim 50\%$. The approximation for $\Delta c_\mathrm{H}$ is more accurate, with a relative error of $\sim 5\%$ in the poloidal case [(c) panel] and $\sim 7\%$ in the toroidal case [(d) panel]. In any case, we see no dependence of the error on the compactness of the EoS. As for $\Delta c_\mathrm{s}$, its approximation in Fig.~\ref{fig:pca_cs_stt} holds with a relative error mostly under $\sim 10\%$. We note that the coefficients for $\Delta c^\mathrm{PCA}_\mathrm{H}$ in Eq.~\ref{eq:ch_pca_stt} in the toroidal case can be used also in the poloidal case, but in this case the 90th percentile relative error increases to $\sim 30\%$. However, similarly to what we found in the GR case, performing $c^\mathrm{PCA}_\mathrm{H} \rightarrow 3/2c^\mathrm{PCA}_\mathrm{H}-0.2$ allows one to use the toroidal coefficients in the poloidal case with a $\sim 10\%$ error.
\\\\
If we apply Eq.~\ref{eq:cb_pca_stt} to approximate $\Delta c_\mathrm{B}$ in models computed using the POL2 EoS, errors remain roughly the same in the poloidal case, while they increase by $\sim 20\%$ in the toroidal case. Instead, $\Delta c_\mathrm{H}$ approximated using Eq.~\ref{eq:ch_pca_stt} holds also for the POL2 EoS, at the expense of an error reaching $\sim 15\% (\sim 30\%)$ in the poloidal (toroidal) case. If we approximate $\Delta c_\mathrm{s}$ using Eq.~\ref{eq:cs_pca_stt} for models computed with the POL2 EoS, the PCA approximation presents a deviation mostly under $\sim 40\%$.
\\\\
We note here that the deformation coefficient, $c_\mathrm{B}$, computed for STT models at the same central density is approximately the same for any value of $\beta _0 \in \{ -6,-5.75,-5.5,-5 \}$; this happens both for purely poloidal and purely toroidal magnetic configurations. This suggests that it is ultimately the central density that determines the deformation coefficient $c_\mathrm{B}$ of a NS: the role of the scalar field is that of merely shifting the central density of a model with the same mass to different values with respect to GR. For the same reason, models for STTs computed at a fixed Komar mass have different $c_\mathrm{B}$: their central density varies with $\beta _0$.
\\\\
A time-varying quadrupolar deformation leads to the emission of GWs. While in GR these are only of tensor nature (i.e. the wave carrier is a spin-2 particle), in the case of a STT, a scalar channel is also present (whose energy is carried by a spin-0 particle). While the multipolar pattern of the energy carried by tensor GWs cannot contain lower-than-quadrupole modes, scalar GWs can contain any multipolar component. A time-dependent monopolar variation in the structure of a scalarised NS leads to the emission of monopolar waves; being that monopoles rotationally invariant, this cannot happen simply due to the rotation of the NS, but there must also be some kind of radial time-dependent variation, for instance, when the star collapses \citep{gerosa_numerical_2016}]. For this reason, we only focus on quadrupolar modes of GWs, both tensor and scalar in nature. Given that NSs for GR and for STTs can posses rather different quadrupolar deformations (see e.g. Figs.~\ref{fig:pca_cbh_stt}-\ref{fig:pca_cs_stt}), it is interesting to compare the amount of GWs emitted by NS in these two modes. For this reason, we introduce the following ratio:
\begin{equation}\label{eq:sratio}
        \mathcal{S}= \bigg | \frac{q_\mathrm{s}}{q_\mathrm{g}} \bigg |,
\end{equation}
where
\begin{align}
        &q_\mathrm{s} = 2\pi \int \alpha _\mathrm{s} \mathcal{A}^4 T_\mathrm{p} \left( 3\sin ^2 \theta -2 \right) r^4 \sin \theta \mathrm{d}r \mathrm{d}\theta , \\
        &q_\mathrm{g} = \int \left[ \pi \mathcal{A}^4 (\varepsilon+\rho)-\frac{1}{8} \partial \chi \partial \chi \right] r^4 \sin \theta \left(3\sin^2\theta -2 \right)  \mathrm{d}r \mathrm{d}\theta ,
\end{align}
are, respectively, Newtonian approximations of the `trace quadrupole' and of the `mass quadrupole' of the NS. The mass quadrupole $q_\mathrm{g}$ is $I_\mathrm{zz}-I_\mathrm{xx}=e I_\mathrm{zz}$ (see Eq.~\ref{eq:deform}) and acts as the source of tensor waves. The trace quadrupole is related to the `quadrupolar deformation of the trace', defined in \citetalias{soldateschi_2020} via Eq.~C.19, which is the source of scalar waves. Thus, the ratio $\mathcal{S}$ computed for a NS model measures in which channel that model will emit most quadrupolar GWs, either the tensor ($\mathcal{S}<1$) or the scalar channel ($\mathcal{S}>1$). We found that the following quasi-universal relations hold for $\mathcal{S}$:
\begin{equation}\label{eq:sratio_pca_stt}
        \mathcal{S}^\mathrm{PCA}=
            \begin{cases}
                1.98^{+0.18}_{-0.05} R_{10}^{-0.71} M_{1.6}^{-0.54} Q_{1}^{1.22} \;{\rm for\; poloidal,}\\\\
                1.99^{+0.18}_{-0.07} R_{10}^{-0.74} M_{1.6}^{-0.60} Q_{1}^{1.23} \;{\rm for\; toroidal.}
            \end{cases}
\end{equation}
\begin{figure*}[!ht]
  \centering
  \subfloat[][]{\includegraphics[width=.49\textwidth]{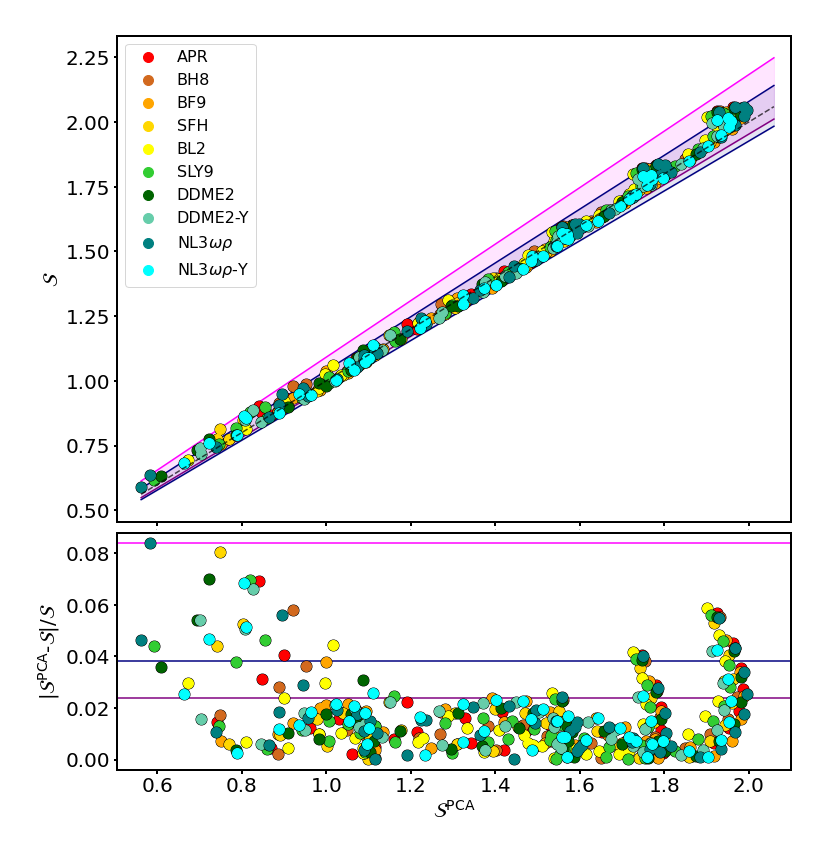}}\quad
  \subfloat[][]{\includegraphics[width=.49\textwidth]{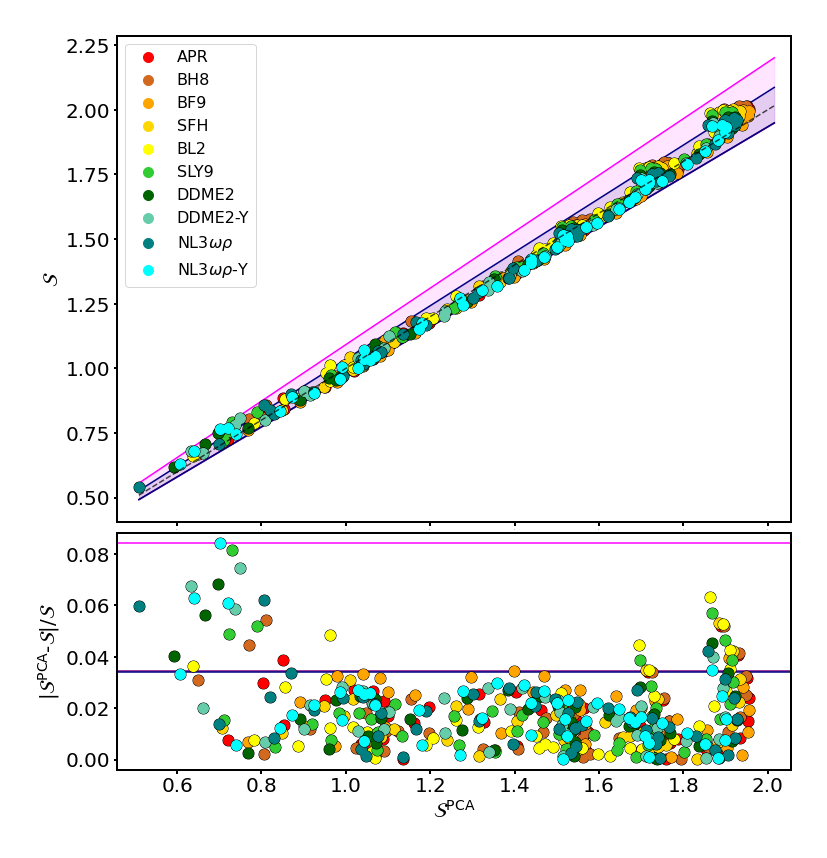}}
  \caption{Ratio $\mathcal{S}$ between scalar and tensor quadrupolar GW losses, calculated according to Eq.~\ref{eq:sratio} for STTs with $\beta _0~\in~\{ -6,-5.75,-5.5,-5 \}$. This is plotted versus its approximation $\mathcal{S}^\mathrm{PCA}$, calculated with the quasi-universal relations in Eq.~\ref{eq:sratio_pca_stt} (top plot in each panel). The corresponding relative devitations from the PCA are given in the bottom plot in each panel. Panel (a) refers to a purely poloidal magnetic field; panel (b) refers to a purely toroidal magnetic field. The dashed line is $\mathcal{S} = \mathcal{S}^\mathrm{PCA}$. The magenta shaded area comprises all data points and the purple and magenta lines represent the upper and lower bounds of Eq.~\ref{eq:sratio_pca_stt}. The dark blue lines bounding the shaded blue area mark the 90th percentile error region. The EoS are colour-coded, and ordered in the legend according to the compactness $C=M_\mathrm{k}/R_\mathrm{c}$ calculated at $M_\mathrm{k}=1.4$M$_\odot$ in GR: red for the highest compactness and blue for the lowest compactness.}
  \label{fig:pca_sratio_stt}
\end{figure*}
The values of $\mathcal{S}$, computed through Eq.~\ref{eq:sratio} for $\beta _0~\in~\{-6,-5.75,-5.5,-5\}$, are plotted in Fig.~\ref{fig:pca_sratio_stt}, against their PCA approximation, computed via Eq.~\ref{eq:sratio_pca_stt}. We see that the approximation is quite accurate in both the poloidal and the toroidal case, with a maximum relative error of $\sim 8\%$ in both cases, but mostly concentrated under $\sim 4\%$. We see that the coefficients in Eq.~\ref{eq:sratio_pca_stt} are practically identical in the poloidal and the toroidal case; in fact, using the coefficients of the PCA approximation of the toroidal case in the poloidal case leads to an only slightly larger error (around $\sim 1\%$ higher). As we discuss in Sect.~\ref{sec:constr}, we believe that this similarity points to the existence of a relation between the mass and trace quadrupoles that does not depend on either the magnetic field geometry or the EoS.

\subsection{Equations of state containing strange quark matter}\label{subsec:sqm}
As we see in Fig.~\ref{fig:cbh_mk_gr}, the EoS describing strange quark stars exhibit quite different behaviours in the distortion coefficients than the other EoS, especially with regard to the SQM2 EoS, and even more so in the case of $c_\mathrm{H}$ in the toroidal case (b panel, bottom plot). Overall, we see that stars described by the SQM2 EoS are more deformable than all of the other EoS, when compared at the same mass. The reason for this is easily seen from the mass-radius relation in Fig.~\ref{fig:mrrel}: at the same mass, strange quark stars described by the SQM2 EoS possess a larger radius, and are thus less compact and more prone to deformation. Instead, the SQM1 EoS, for masses lower than $\sim 1.3$M$_\odot$, is the most compact one; however, we note that while the compactness of an EoS is indeed related to its deformability as defined in Eq.~\ref{eq:deform}, it is not the only ingredient: the interplay of the internal magnetic field of the star and its density distribution is another effect that plays an important role on how the magnetic field deforms the star structure. Indeed, strange quark stars possess a much `flatter' density distribution than standard NSs, in which the density gradient is larger. As we anticipated in Sect.~\ref{sec:results}, toroidal models described by the SQM1 and SQM2 EoS show a discontinuity in the magnetic field profile at the surface; since this behaviour leads to a non-vanishing Lorentz force, we caution the reader that these models might not be true equilibria, and thus our results in this case might not be completely accurate. Moreover, we note that the definition of the gravitational binding energy $W$ in Eq.~\ref{eq:distcoeff}, being the same for all EoS, does not include the contribution of the QCD vacuum energy: the distortion coefficient, $c_\mathrm{H}$, we find is the one computed ignoring this additional energy, and is roughly a fraction $W/W^\mathrm{QCD}$ different from the `real' coefficient, where $W^\mathrm{QCD}$ is the binding energy which includes the QCD vacuum. The QCD vacuum energy is expected to have a more important contribution to the binding energy in models which possess a low mass, thus possibly explaining the different behaviour of the SQM2 curve in the bottom plot of the (b) panel in Fig.~\ref{fig:cbh_mk_gr}.

Due to the extent of these differences, applying the quasi-universal relations we found for standard NSs to the case of SQM1 and SQM2 leads to greater errors: in GR, $c^\mathrm{PCA}_\mathrm{B}$ is a factor  of $\sim 0.4-0.8$ lower than $c_\mathrm{B}$ for both purely poloidal and toroidal magnetic fields. For $c_\mathrm{H}$, using Eq.~\ref{eq:ch_pca_gr} the maximum error increases to $\sim 8\% (\sim 12\%)$ for purely poloidal (toroidal) magnetic fields in the case of SQM1, and $\sim 5\% (\sim 40\%)$ for purely poloidal (toroidal) magnetic fields in the case of SQM2. As for $c^\mathrm{PCA}_\mathrm{s}$, it is at most a factor of $\sim 1.4$ higher than $c_\mathrm{s}$. For STTs, $\Delta c^\mathrm{PCA}_\mathrm{B}$ in the poloidal case is around a factor of $\sim 2$ lower (higher) than $\Delta c_\mathrm{B}$ for the SQM1 (SQM2) EoS, excluding a few outlier points; in the toroidal case, it is a factor of $\sim 2$ lower for both SQM1 and SQM2. For $\Delta c_\mathrm{H}$, using Eq.~\ref{eq:ch_pca_stt} the maximum error increases to $\sim 30\% (\sim 50\%)$ for purely poloidal (toroidal) magnetic fields, again ignoring few outlier points. In the case of $\Delta c^\mathrm{PCA}_\mathrm{s}$, it is around a factor  of $\sim 1.7$ lower than $\Delta c_\mathrm{s}$.

\section{EoS and magnetic structure constraints}\label{sec:constr}
\begin{figure}[!ht]
  \centering
  \includegraphics[width=\columnwidth]{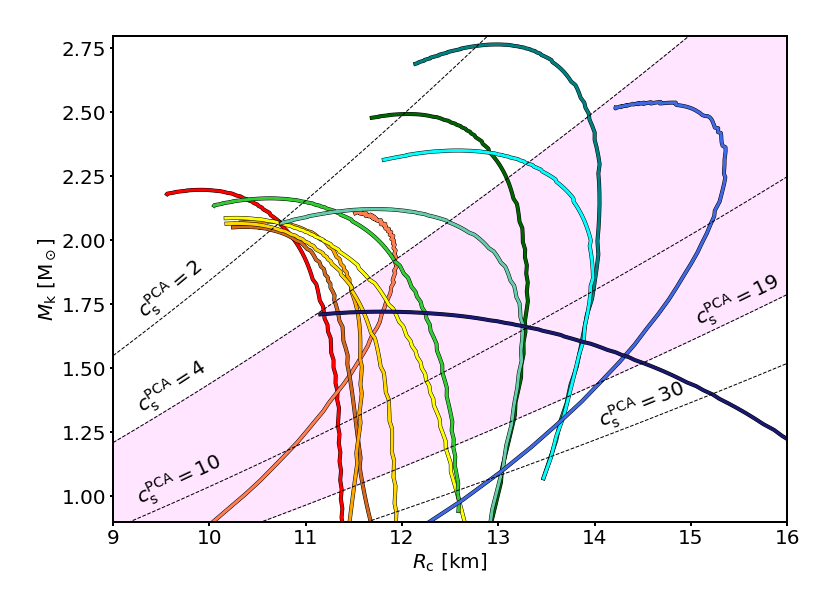}
  \caption{Komar mass $M_\mathrm{k}$ against circumferential radius $R_\mathrm{c}$ for un-magnetised, static models of NSs computed with the EoS described in Sect.~\ref{sec:eos} in GR. The EoS are colour-coded, and ordered in the legend, according to the compactness $C=M_\mathrm{k}/R_\mathrm{c}$ calculated at $M_\mathrm{k}=1.4$M$_\odot$ in GR: red for the highest compactness and blue for the lowest compactness. The black dashed lines denote $c^\mathrm{PCA}_\mathrm{s}=\mathrm{constant}$ isolines. The two values $c^\mathrm{PCA}_\mathrm{s} = 4-19$ containing the magenta highlighted area are those found in the case of the example in Sect.~\ref{sec:conclusions}.}
  \label{fig:cs_isosurfaces}
\end{figure}

As argued in Sect.~\ref{sec:intro}, among the major uncertainties in NS physics are the EoS and the magnetic field structure of their inner regions. The quasi-universal relations we found may be useful in this sense, given that they are independent of the EoS of standard NSs, thus leaving the NS internal magnetic structure as the only major unknown in GR. The most promising relation is that of $c_\mathrm{s}$. On the one hand, $c_\mathrm{s}$ can be computed from its definition Eq.~\ref{eq:distcoeff_cs} if one is able to measure both the magnetic field strength at the surface of the NS, $B_\mathrm{s}$, and its quadrupolar deformation $e$. The latter is in turn inferable from the strain of CGWs emitted by the NS, $h_0\propto eI$, where $I$ is the moment of inertia of the NS along its rotation axis, which must be unaligned to its magnetic axis. The moment of inertia, $I,$ is a function of the NS mass and radius, and, in principle, it depends on the EoS. However, \citet{breu_2016}\footnote{We note that the moment of inertia $I$ used in \citet{breu_2016} is defined as the ratio of the angular momentum to the angular velocity, and is different from the definition of $I_{zz,xx}$ we use in this paper, which is Newtonian. While the two moments of inertia can differ, it was found that the quadrupolar deformation $e$ defined as in Eq.~\ref{eq:deform}, being a ratio, is very similar in both regimes \citep{pili_general_2015}.} have found that an EoS-independent relation between $I$, the NS mass and its radius exists. On the other hand, Eq.~\ref{eq:cs_pca_gr} allows us to estimate $c^\mathrm{PCA}_\mathrm{s}$ by knowing just the NS mass and radius (see Fig.~\ref{fig:cs_isosurfaces}, where a few $c^\mathrm{PCA}_\mathrm{s} = \mathrm{const.}$ isolines are plotted). Thus, three possible informative scenarios may arise:
\begin{enumerate}
    \item $c_\mathrm{s}<c^\mathrm{PCA}_\mathrm{s}$ and $I$ is computed with the EoS-independent relation by \citet{breu_2016}: the only assumption made is that of a very specific purely poloidal magnetic field permeating the NS. In this case, the deformation coefficient is reduced by the presence of a toroidal component, which counteracts the deformation of the poloidal one. The strength of the toroidal component increases with respect to the poloidal one the larger $c_\mathrm{s}$ deviates from $c^\mathrm{PCA}_\mathrm{s}$.
    \item $c_\mathrm{s}>c^\mathrm{PCA}_\mathrm{s}$ and $I$ is computed with the EoS-independent relation by \citet{breu_2016}: Since the star is more deformed than what the extremal case of a purely poloidal field can produce, there must be another source of deformation other than the magnetic field .
    \item $c_\mathrm{s}>c^\mathrm{PCA}_\mathrm{s}$ and $I$ is computed by assuming an EoS: either there is another source of deformation, other than the magnetic field, or the assumed EoS in not consistent, because it predicts a moment of inertia, $I,$ that is not compatible with the deformation coefficient of the star.
\end{enumerate}
If $c_\mathrm{s}<c^\mathrm{PCA}_\mathrm{s}$ and $I$ is computed by assuming an EoS, not much can be said without further information because both the magnetic field geometry and the EoS are assumed. We stress that our analysis, and thus the three informative scenarios just described, are valid in the case of a purely poloidal magnetic configuration that satisfies the criterion for equilibrium in the Bernoulli formalism: different magnetic configurations should be explored to possibly strengthen our conclusions, even if it is not clear whether purely poloidal equilibria that significantly differ from the one we have adopted do exist. To recap, once the quantities $M_\mathrm{k},R_\mathrm{c}$ of a NS are measured, a point in the mass-radius diagram Fig.~\ref{fig:cs_isosurfaces} can be placed. On the one hand, its deformation coefficient $c_\mathrm{s}$ can be computed if one is able to measure also the quantities $h_0,B_\mathrm{s}$; on the other hand, $c^\mathrm{PCA}_\mathrm{s}$ is computed from Eq.~\ref{eq:cs_pca_gr}, corresponding to a certain isoline in Fig.~\ref{fig:cs_isosurfaces}. Depending on whether $c_\mathrm{s}<c^\mathrm{PCA}_\mathrm{s}$ or $c_\mathrm{s}>c^\mathrm{PCA}_\mathrm{s}$ and on whether one has computed the NS moment of inertia $I$ through the EoS-independent relation of \citet{breu_2016} or by assuming an EoS, various conclusions on the NS internal magnetic structure or on the consistency of the EoS can be asserted.
\\\\
Similar conclusions can be drawn if one considers the other two deformation coefficients, $c_\mathrm{B}$ and $c_\mathrm{H}$, instead of $c_\mathrm{s}$, with the caveat that these quantities are defined also for purely toroidal magnetic fields and require one to be able to measure the maximum strength of the magnetic field, $B_\mathrm{max}$, or the ratio $\mathcal{H}/W$, respectively. However, such quantities are much less likely to be measured than the magnetic field at the surface.
For this reason, quasi-universal relations Eqs.~\ref{eq:cb_pca_gr}-\ref{eq:ch_pca_gr} may be more useful to constrain $B_\mathrm{max}$ or $\mathcal{H}/W$ themselves. More specifically, $B_\mathrm{max} \approx (e/c_\mathrm{B})^{1/2}$ and $\mathcal{H}/W \approx e/c_\mathrm{H}$; given that $c_\mathrm{B,H}<c^\mathrm{PCA}_\mathrm{B,H}$, once the NS mass and radius are known, it is possible to estimate a lower bound for both $B_\mathrm{max}$ and $\mathcal{H}/W$ by using the quasi-universal relations found in the purely poloidal and purely toroidal case. We note that the quasi-universal relations for $c_\mathrm{B}$ and $c_\mathrm{H}$, as well as that of $c_\mathrm{s}$, are in any case useful for the purpose on numerical simulations and theoretical estimations, allowing us to quickly and easily determine the distortion coefficients from the mass and radius of a model -- without going through a full numerical simulation.
\\\\
In STTs, the scalar charge is also unkown. As discussed in Sect.~\ref{sec:intro}, some of the effects of changing the NS EoS are degenerate with the presence of a non-negligible scalar charge. For this reason, the quasi-universal relations in Eqs.~\ref{eq:cb_pca_stt}-\ref{eq:ch_pca_stt}-\ref{eq:cs_pca_stt} may help us to understand whether a distortion coefficient inferred from observations, via the relations in Eqs.~\ref{eq:distcoeff}-\ref{eq:distcoeff_cs}, is compatible with a non-zero scalar charge of the observed NS, independently from its EoS. To this end, the relation in Eq.~\ref{eq:sratio_pca_stt} may prove to be more promising because it does not require any knowledge of the strength of the magnetic field. In particular, since Eq.~\ref{eq:sratio_pca_stt} is essentially the same in both poloidal and toroidal magnetic configurations, it is possible that $\mathcal{S}$ is independent of the magnetic field altogether. In this case $\mathcal{S} \approx \mathcal{S}^\mathrm{PCA}$, and the observation of CGWs from a given source of known mass, radius, distance $d$ and spin period $P$ translates into a lower bound for a function of the scalar charge: $g(Q_\mathrm{s},q_\mathrm{s}) < f(M_\mathrm{k},R_\mathrm{c},d,P,h_0^\mathrm{min})$, where $h_0^\mathrm{min}$ is the sensitivity of a given GW detector at the frequency $2/P$ and $g(Q_\mathrm{s},q_\mathrm{s})$ is zero for a non-scalarised NS. On the other hand, the non-observation of CGWs for a given NS can be translated into an upper bound: $g(Q_\mathrm{s},q_\mathrm{s}) > f(M_\mathrm{k},R_\mathrm{c},d,P,h_0^\mathrm{min})$.
\\\\
Next, we comment on how our results compare to the previous findings of \citet{cutler_2002}. There, the author found that in Newtonian theory, for an incompressible, constant-density NS, the distortion coefficient $c_\mathrm{H}=15/4\sim 3.75$ in the case of a purely toroidal magnetic field, while it is $c_\mathrm{H}\sim 15/2\sim 7.50$ for a purely poloidal model. As we showed in Fig.~\ref{fig:cbh_mk_gr}, we found that $c_\mathrm{H}$ ranges from $\sim 2$ to $\sim 3$ (from $\sim 2.5$ to $\sim 4$) for toroidal (poloidal) models. Therefore, by using the constant coefficients of \citet{cutler_2002}, found under some simplifying assumptions, one would find a $c_\mathrm{H}$ that is a factor of$~\sim 1.25-1.88 (\sim 1.88-3.00)$ higher than our estimates; {this could in part be the result of having different magnetic field geometries, in turn stemming from different assumptions on the NS structure}.  Finally, as explained in \citet{frieben_equilibrium_2012}, when considering a NS with a superconducting core, one expects to find an increase in $c_\mathrm{H}$, with respect to a non-superconducting star, which is roughly a factor of $\langle BB_\mathrm{c1}\rangle /\langle B^2 \rangle$, where $B$ is the magnitude of the magnetic field, $B_\mathrm{c1}\approx 10^{15}$G is the first critical magnetic field strength and $\langle \dots \rangle$ indicates volume average. This is valid only as a first approximation for $\langle B \rangle <10^{15}$G. Thus, in this limit, for superconducting models, we expect to find distortion coefficients that are roughly a factor of $B_\mathrm{c1}/\langle B \rangle$ higher than what we found.

\section{Conclusions}\label{sec:conclusions}
In this work, we explore the relation among the magnetic deformability and the main observable quantities of NS models described by a variety of different EoS allowed by observational and nuclear physics constraints. We did so in the case of static, axisymmetric configurations endowed by specific choices of either purely poloidal or purely toroidal magnetic fields based upon two different theories of gravitation: GR and a massless STT containing the spontaneous scalarisation phenomenon. We used 12 different EoS which satisfy the latest astrophysical and nuclear physics constraints, plus a polytropic law that has been widely used in the literature. These EoS span a wide range of calculation methods and particle contents, ranging from zero-temperature to finite-temperature ones, from nucleonic EoS to those containing hyperons or quark degreees of freedom; moreover, we considered two EoS that are capable of describing strange quark stars.

We first obtained the distortion coefficients which describe the magnetic deformation of NSs in the limit of non-extreme magnetic fields, that is, when the quadrupolar magnetic deformation of the star follows a quadratic law with the magnetic field magnitude. In particular, we studied three different coefficients by parametrising the quadrupolar deformation of our models with their maximum magnetic field strength, their magnetic energy to gravitaitonal binding energy ratio, and their superficial magnetic field strength: $c_\mathrm{B}$, $c_\mathrm{H}$, and $c_\mathrm{s}$ respectively. While the first two may be most useful for the purpose of computing the deformation of NS models without going through a full numerical simulation, the latter may be of help in constraining the magnetic properties in NS interiors. We find that while $c_\mathrm{H}$ varies by a maximum factor of 1.5 among all the models we studied, $c_\mathrm{B}$ and $c_\mathrm{s}$ exhibit a much stronger dependence on the NS mass. Moreover, the behaviour of $c_\mathrm{B,H}$ is qualitatively similar for poloidal and toroidal configurations. Since the polytropic EoS POL2 and the strange quark matter EoS SQM1 and SQM2 exhibit a radically different behaviour with respect to all other EoS, we focus mainly on the ten EoS - allowed by  the observations - describing standard NSs.

Subsequently, we looked for relations among the three coefficients, the NS Komar mass, and their circumferential radius; in the case of STTs, we also consider the dependence on their scalar charge. Specifically, we are most interested in EoS-independent relations, or quasi-universal relations. We find that there are equations at hand to describe the distortion coefficients in term of the NS mass, radius, and scalar charge (in STTs) that are valid for all the ten standard EoS to a satisfying level of accuracy. These relations have a simple form, consisting only of power laws or polynomials. We find that these relations can be applied also to the case of the POL2, SQM1, and SQM2 EoS, but with a (sometimes significantly) reduced accuracy. This is due to a number of reasons: the POL2 EoS is a simple polytrope, which we include only for reference, and it lacks all the facets of various particle contents that are described by the other EoS; the SQM1 and SQM2, instead, describe a different type of star, which is predicted to exist - in which case it would help to solve a number of problems regarding NS astrophysics - but one that possesses a radically different structure than a standard NS. In particular, the density profile throughout quark stars is nearly flat, with a discontinuity at the surface: this leads, in some cases, to a discontinuity in the toroidal magnetic field at the star surface, thus rendering these models not accurate in terms of a true equilibria. Moreover, strange quark stars are able to sustain very large masses with large radii, leading to a magnetic deformation that is, in general, higher than that of standard NSs. Finally, we do not consider the contribution of the QCD vacuum energy to the binding energy of our models.

In the case of STTs, we find other quasi-universal relations linking the mass, radius, and scalar charge of the NS to the ratio of the scalar-to-tensor GWs strain, $\mathcal{S}$. {We find that $\mathcal{S}$ is, within the numerical accuracy of our code, well approximated by a single expression, $\mathcal{S}^\mathrm{PCA}$, both in the purely poloidal and purely toroidal cases. This suggests that $\mathcal{S}$ may be a quantity which is independent of the magnetic field altogether, although this point should be better addressed by simulations involving NS models endowed with mixed fields.}
\\\\
The quasi-universal relations we find depend on potentially observable quantities. In particular, by knowing the NS mass and radius, it is possible to directly compute its distortion coefficients as predicted by the relations we found, which are valid in the case of purely poloidal or purely toroidal magnetic fields. By comparing such values with those we computed through the definitions Eqs.~\ref{eq:distcoeff}-\ref{eq:distcoeff_cs}, we can infer information on the magnetic structure hidden in the NS interior. This is possible because the pure magnetic configurations we consider are extremal, in the sense that purely poloidal and purely toroidal magnetic fields act on the quadrupolar deformation of the NS in an opposite way; as such, a pure configuration exerts the maximum deformation on a NS, {while deviations from this case may imply a different magnetic configuration than the one we assume}. Since only the superficial magnetic field is accessible through direct observations, we expect the relation regarding $c_\mathrm{s}$, Eq.~\ref{eq:cs_pca_gr}, to be the most useful in this sense. However, Eqs.~\ref{eq:cb_pca_gr}-\ref{eq:ch_pca_gr} may also show their utility for computing the distortion coefficients of a NS model by knowing only its mass and radius, without going through a full numerical simulation. We stress that only specific choices of purely poloidal and toroidal magnetic configurations have been investigated, and different choices of the magnetisation functions should be explored to possibly strengthen our conclusions.
\\\\
The relation involving $\mathcal{S}$, namely, Eq.~\ref{eq:sratio_pca_stt}, along with the considerations we described regarding pure magnetic configurations, translates into a lower bound on the scalar charge of a NS whose mass, radius, and quadrupolar deformation are known. For this reason, it may be useful in constraining the theory of gravity.
\\\\
We note that in order to obtain the quadrupolar deformation of an isolated NS, it is necessary to detect CGWs emitted by them. While this is within the scope of current GW observations [see e.g. \citet{abbott_cgw_2020}, where it is suggested that the particular spin-down and glitch behaviour of pulsar PSR J0537-6910 ought to be attributed to the emission of CGWs], no signature has been found yet. We can use the quasi-universal relations we found to assess the detectability of known NSs. For example, using Eq.~\ref{eq:cs_pca_gr} and the quasi-universal relation for $I$ found by \citet{breu_2016}, we can set limits on the minimum surface magnetic field strength $B^\mathrm{min}_\mathrm{s}$ that produces a deformation leading to detectable CGWs. Using the distance $d=0.16$kpc and rotation period $P=5.758$ms of the closest known millisecond pulsar [J0437-4715, see \citet{atnf_2005}], which has a  mass of 1.44M$_\odot$, and taking a radius in the range of $10-14$km (corresponding to $c^\mathrm{PCA}_\mathrm{s} = 4-19$ respectively, see Fig.~\ref{fig:cs_isosurfaces}), we find $B^\mathrm{min}_\mathrm{s}=1.4\times 10^{13}-4.1\times 10^{13}$G for a 2 year long observation with the advanced LIGO detector (aLIGO), $B^\mathrm{min}_\mathrm{s}=8.6\times 10^{12}-2.6\times 10^{13}$G for a 1 month long observation with the advanced Einstein Telescope (ET) detector, $B^\mathrm{min}_\mathrm{s}=4.3\times 10^{12}-1.3\times 10^{13}$G for a two-year long observation with ET.

When considering NSs endowed with a superconducting core, as previously explained, the effective magnetic field entering the distortion coefficients increases, which reflects in a lower $B^\mathrm{min}_\mathrm{s}$: $B^\mathrm{min}_\mathrm{s}=1.8\times 10^{11}-1.6\times 10^{12}$G for a two-year long observation with aLIGO, $B^\mathrm{min}_\mathrm{s}=7.4\times 10^{10}-6.6\times 10^{11}$G for a one-month long observation with ET, $B^\mathrm{min}_\mathrm{s}=1.8\times 10^{10}-1.6\times 10^{11}$G for a two-year long observation with ET. We note that these values are found in the best case scenario, namely, when the magnetic and the rotation axes are orthogonal and the latter points towards the observer. These values, while they are not unrealistically high, they are, in the best case, about one order of magnitude larger than the average surface dipole magnetic fields measured in millisecond pulsars \citep{cruces_2019}. However, such measured values may be low because of a variety of reasons, ranging from screening due to accreted matter \citep{romani_1990} to ambipolar diffusion \citep{cruces_2019}. Since the magnetic field producing the distortion in Eq.~\ref{eq:distcoeff_cs} is the one below any accreted material, namely, the un-screened one, the possibility of the magnetic deformation producing CGWs that would be detectable by future GW detectors may be more promising. In the case of millisecond magnetars, the surface magnetic field is expected to be on the order of $\sim 10^{14-15}$G \citep{dallosso_2021}, which would significantly enhance the possibility of their detection by CGWs emissions.
\\\\
In this paper, we focus only on the quadrupolar deformation of NSs due to either purely poloidal or purely toroidal magnetic fields. While we expect that mixed-fields configurations, such as the twisted-torus one, yield deformations contained between the limiting values we found in the pure cases we studied {(and, thus, distortion coefficients which are smaller than those we found)}, this point should be better addressed by numerical simulations containing mixed-fields geometries. {We stress that although only very special configurations have been investigated and, therefore, the results we find can probably only give order-of-magnitude estimates for other magnetic
field structures, the dependence on stellar parameters might be universal, and this aspect could be investigated with other examples.} We recall that the only known formalism to compute equilibria in the full non-linear regime is through the approach we use, that is, through the use of the generalised Bernoulli integral, which sets severe constraints on the possible magnetic field geometry. As an example, poloidal configurations are dominated by the dipole term, but also contain higher order multipoles. It has been found \citep{mastrano_2013} that higher order multipoles can contribute to the magnetic energy even more than the dipole field, consequently increasing the deformation and, thus, the detectability of these systems. Moreover, it is expected that at least some NSs contain also higher-multipole magnetic fields, due to a mismatch between the surface magnetic field strength obtained by observed X-ray spectra and the inferred dipole field \citep{guver_2011}. It would also be interesting to understand whether a quasi-universal relation like those we find exists also for the quadrupole deformation $e$, without expanding it in terms of the distortion coefficients.
\\\\
Finally, while we opted for the PCA algorithm to search for simple relations between the physical quantities of interest, it would be interesting to use a different technique, for instance, autoencoders; with the main difference being that the former looks for linear relations in the data, while the latter is a generalisation to non-linear maps. This means that while we must adopt some analytical form of the relation we wish the PCA to find, more freedom is allowed via the application of  autoencoders.

\begin{acknowledgements}
The authors acknowledge financial support from the Accordo Attuativo ASI-INAF n. 2017-14-H.0 `On the escape of cosmic rays and their impact on the background plasma', the PRIN-INAF 2019 `Short gamma-ray burst jets from binary neutron star mergers' and from the INFN Teongrav collaboration.
\end{acknowledgements}



\bibliographystyle{aa}
\interlinepenalty=10000
\bibliography{articolo} 


%
%
%


\end{document}